\documentclass[journal,onecolumn,final,twoside,letterpaper,11pt]{IEEEtran}

\usepackage{subfigure,cite,graphicx,amsmath,amssymb,eufrak,mathrsfs,epsf,epsfig,dblfloatfix}
\newtheorem{theorem}{Theorem}
\newtheorem{lemma}{Lemma}

\newtheorem{corollary}{Corollary}
\newtheorem{remark}{Remark}

\begin{document}
\title{Degrees of Freedom Region of a Class of Multi-source Gaussian Relay Networks}

\author{Sang-Woon Jeon, \emph{Student Member}, \emph{IEEE}, Sae-Young Chung, \emph{Senior Member}, \emph{IEEE}, and\\
Syed A. Jafar, \emph{Senior Member}, \emph{IEEE}\\
\thanks{The work of S.-W. Jeon and S.-Y. Chung was supported by the MKE under the ITRC program supervised by the NIPA (NIPA-2010-(C1090-1011-0011)).
The work of Syed A. Jafar was supported by NSF under grants CCF-0830809 and CCF-0963925 and by ONR under grant N00014-08-1-0872.
The material in this paper was presented in part at the Allerton Conference on Communication, Control, and Computing, Monticello, IL, USA, September 2009 and at the IEEE Information Theory Workshop, Dublin, Ireland, August/September 2010.}
\thanks{S.-W. Jeon and S.-Y. Chung are with the Department of Electrical Engineering, KAIST, Daejeon, South Korea (e-mail: swjeon@kaist.ac.kr; sychung@ee.kaist.ac.kr).}
\thanks{S. A. Jafar is with the Electrical Engineering and Computer Science, University of California, Irvine, CA 92697, USA (e-mail: syed@uci.edu).}
}



\maketitle

\begin{abstract}
We study a layered $K$-user $M$-hop Gaussian relay network consisting of $K_m$ nodes in the $m^{\operatorname{th}}$ layer, where $M\geq2$ and $K=K_1=K_{M+1}$.
We observe that the time-varying nature of wireless channels or fading can be exploited to mitigate the inter-user interference.
The proposed amplify-and-forward relaying scheme exploits such channel variations and works for a wide class of channel distributions including Rayleigh fading.
We show a general achievable degrees of freedom (DoF) region for this class of Gaussian relay networks.
Specifically, the set of all $(d_1,\cdots, d_K)$ such that $d_i\leq 1$ for all $i$ and $\sum_{i=1}^K d_i\leq K_{\Sigma}$ is achievable, where $d_i$ is the DoF of the $i^{\operatorname{th}}$ source--destination pair and $K_{\Sigma}$ is the maximum integer such that $K_{\Sigma}\leq \min_m\{K_m\}$ and $M/K_{\Sigma}$ is an integer.
We show that surprisingly the achievable DoF region coincides with the cut-set outer bound if $M/\min_m\{K_m\}$ is an integer, thus interference-free communication is possible in terms of DoF.
We further characterize an achievable DoF region assuming multi-antenna nodes and general message set, which again coincides with the cut-set outer bound for a certain class of networks.
\end{abstract}


\section{Introduction} \label{sec:intro}
\IEEEPARstart{C}{haracterizing} the capacity of \emph{Gaussian relay networks} is one of the fundamental problems in network information theory.
However, for Gaussian relay networks, the signal transmitted from a node will be heard by multiple nodes (broadcast) and a node will receive a superposition of the signals transmitted from multiple nodes (interference) and there exist fading and noise, which make the problem complicated.
To overcome such difficulties, simplified wireless network models have been developed in \cite{ArefPD:80,Ray:03, Bhadra:06,Niranjan:06,Smith:07,AvestimehrDiggaviTse:07} that provide intuition towards an approximate capacity characterization of single-source Gaussian relay networks \cite{AvestimehrDiggaviTse:08, Lim:00}.

Unlike the single-source case, the capacity or an approximate capacity characterization of \emph{multi-source Gaussian relay networks} is very challenging since the transmission of other sessions acts as the \emph{inter-user interference}.
Due to the interference, the multi-source extension from the results in \cite{AvestimehrDiggaviTse:08, Lim:00} is not straightforward.
Recently, remarkable progress has been made on multi-source problems in \cite{Bresler:07, Etkin:08, Viveck1:08, Tiangao:09, Viveck2:09} and the references therein.
It was proved in \cite{Etkin:08} that the Han--Kobayashi scheme indeed achieves the capacity of the two-user Gaussian interference channel within one bits/s/Hz.
The capacity of the $K$-user Gaussian interference channel has been characterized in \cite{Viveck1:08} as
\begin{equation}
\frac{K}{2}\log(P)+o(\log(P))
\end{equation}
if channel coefficients are sufficiently independent and drawn from a continuous distribution, where $P$ denotes the signal-to-noise ratio (SNR).
To show the degrees of freedom (DoF) or capacity pre-log term of $K/2$, the technique of \emph{interference alignment} was used, which minimizes the overall interference space by aligning multiple interfering signals from unintended sources at each destination.
The concept of interference alignment has also been used to characterize the DoF of the $K$-user multi-antenna Gaussian interference channel \cite{Tiangao:09} and the $X$-network in which each source has independent messages for all destinations \cite{Viveck2:09}.
Another alignment technique called \emph{ergodic interference alignment} has been proposed in \cite{Nazer:09} showing that, for a broad class of channel distributions, half of the interference-free ergodic capacity is achievable for each user in the $K$-user Gaussian interference channel at any SNR.
Based on the inseparability of parallel interference channels \cite{Sankar:08, Viveck4:09}, the ergodic interference alignment scheme jointly encodes messages over two specific channel instances to align the interference.
A similar concept has been also applied for the finite field case in \cite{Nazer:09,JeonITA:09}.

The interference can not only be aligned, but it can also be \emph{cancelled or partially cancelled} for multi-hop Gaussian relay networks.
Assuming amplify-and-forward (AF) relays, each destination may receive multiple copies of an interfering signal from different paths and potentially these copies can cancel each other through a suitable choice of the amplification factors of relays.
Reference \cite{Mohajer:09} has shown that partial interference cancellation using AF relays achieves the capacity of two-user two-hop Gaussian networks within a constant bit gap in some scenarios.
Also, the interference can be completely removed so that the optimal DoF of $K$ is achievable for $K$-user two-hop Gaussian networks if the number of relays is greater than or equal to $K^2$ \cite{Jeon3:09}.

\begin{figure}[t!]
  \begin{center}
  \scalebox{1}{\includegraphics{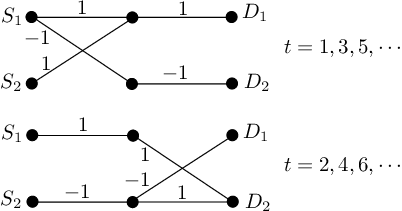}}
  \caption{Example of a two-user two-hop Gaussian relay network, where $S_i$ and $D_i$ denote the $i^{\operatorname{th}}$ source and its destination respectively.}
  \label{fig:example}
  \end{center}
\end{figure}

In this paper, we study layered multi-source multi-hop Gaussian relay networks.
We observe that the \emph{time-varying nature of wireless channels} or \emph{fading} can be exploited to cancel the interference.
As a simple example, consider a two-user two-hop Gaussian relay network in Fig. \ref{fig:example} in which
\begin{equation}
\mathbf{H}_1[t]=\left[
                    \begin{array}{ccc}
                      1 & 1 \\
                      -1 & 0 \\
                    \end{array}
                  \right],
\mathbf{H}_2[t]=\left[
                    \begin{array}{ccc}
                      1 & 0  \\
                      0 & -1 \\
                    \end{array}
                  \right]
\end{equation}
for odd $t$ and
\begin{equation}
\mathbf{H}_1[t]=\left[
                    \begin{array}{ccc}
                      1 & 0  \\
                      0 & -1 \\
                    \end{array}
                  \right],
\mathbf{H}_2[t]=\left[
                    \begin{array}{ccc}
                      0 & -1 \\
                      1 & 1  \\
                    \end{array}
                  \right]
\end{equation}
for even $t$, where $\mathbf{H}_m[t]$ is the $m^{\operatorname{th}}$ hop channel matrix at time $t$.
If odd and even time slots are used separately, each source--destination (S--D) pair can only achieve $1/2$ DoF since there is no path between the first S--D pair for even $t$ and the second S--D pair for odd $t$.
On the other hand, if the relays amplify and forward their signals with one symbol delay, then the interference can be completely cancelled since $\mathbf{H}_2[t+1]\mathbf{H}_1[t]$ becomes the identity matrix.
Hence every S--D pair can achieve one DoF simultaneously.
We generalize this idea to multi-source multi-hop Gaussian relay networks for a wide class of channel distributions including Rayleigh fading.
The key ingredient is to set appropriate delays in AF relaying at each layer such that overall channel matrices become diagonal matrices with non-zero diagonal elements, which guarantees interference-free communication.
Under this class of channel distributions, we show an achievable DoF region of multi-source multi-hop Gaussian relay networks, which characterizes the optimal DoF region if a certain condition is satisfied.
This improves upon our previous result that showed a total of $K$ DoF is achievable for $K$-user $K$-hop networks with $K$ relays in each layer when $K$ is even and a similar technique has been proposed for linear finite field multi-hop networks (see the conference papers \cite{Jeon3:09,JeonISIT2:09}).
We further characterize an achievable DoF region of multi-source multi-hop Gaussian relay networks with multi-antenna nodes and general message set, which is optimal for a certain class of networks.

This paper is organized as follows.
In Section \ref{sec:sys_model}, we explain the underlying system model and define the DoF region.
In Section \ref{sec:result}, we state the main result of this paper, the DoF region of Gaussian relay networks.
In Section \ref{sec:dof_K_K}, we propose an AF relaying scheme and derive its achievable DoF region for $K$-user $K$-hop Gaussian relay networks.
In Section \ref{sec:general_dof}, we generalize this result to $K$-user $M$-hop Gaussian relay networks and show that it characterizes the optimal DoF region if a certain condition is satisfied.
We conclude this paper in Section \ref{sec:conclusion} and refer to Appendix I for the proof of the technical lemma and Appendix II for the proof of the result in Section \ref{sec:dof_K_K} in which $K$ is odd.

\section{System Model} \label{sec:sys_model}
In this section, we explain our network model and introduce encoding, relaying, and decoding functions.
Based on this model, we define the capacity region and the DoF region.
Throughout the paper, we will use $\mathbf{A}$, $\mathbf{a}$, and $\mathcal{A}$ to denote a matrix, vector, and set, respectively.
Let $\prod_{i=1}^K \mathbf{A}_i$ denote $\mathbf{A}_K\mathbf{A}_{K-1}\cdots \mathbf{A}_1$.
The notations used in the paper are summarized in Table \ref{Table:symbols}.

\begin{table}
\caption{Summary of notations} \label{Table:symbols}
\vspace{-0.1in}
\begin{equation*}
    \begin{array}{|c|c|}
    \hline
    \mathbf{A}^{T} (\mbox{ or }\mathbf{a}^{T})& \mbox{Transpose of }\mathbf{A}(\mbox{ or } \mathbf{a})\\
    \hline
    \mathbf{A}^{\dagger} (\mbox{ or }\mathbf{a}^{\dagger})  & \mbox{Conjugate transpose of }\mathbf{A}(\mbox{ or } \mathbf{a})\\
    \hline
    \|\mathbf{A}\|_F (\mbox{ or }\|\mathbf{a}\|)& \mbox{Frobenius norm of }\mathbf{A}(\mbox{ or } \mathbf{a})\\
    \hline
    \operatorname{tr}(\mathbf{A}) & \mbox{Trace of }\mathbf{A}\\
    \hline
    [\mathbf{A}]_{ij} & \mbox{ $(i,j)^{\operatorname{th}}$ element of } \mathbf{A}\\
    \hline
    [\mathbf{A}]_i & \mbox{ $i^{\operatorname{th}}$ row vector of } \mathbf{A}\\
    \hline
    \operatorname{diag}(a_1,\cdots,a_n) & \mbox{Diagonal matrix satisfying }\\
    &  [\operatorname{diag}(a_1,\cdots,a_n)]_{ii}=a_i\\
    \hline
    \mathbf{I}_n & n\times n\mbox{ identity matrix }\\
    \hline
    \mathbf{0}_{n\times m} & n\times m\mbox{ all-zero matrix }\\
    \hline
    \operatorname{real}(a)(\mbox{ or }\operatorname{imag}(a))& \mbox{Real (or imaginary) part of } a\\
    \hline
    |a| & \mbox{Absolute value of } a\\
    \hline
    a^* & \mbox{Complex conjugate of } a\\
    \hline
    \lfloor a\rfloor & \mbox{Floor of } a{~}(\lfloor a\rfloor=\max\{x|x\leq a, x\in\mathbb{Z}\})\\
    \hline
    \operatorname{card}(\mathcal{A})&\mbox{ Cardinality of } \mathcal{A}\\
    \hline
    \mathcal{A}\times \mathcal{B} &\mbox{ Cartesian product of $\mathcal{A}$ and $\mathcal{B}$}\\
    \hline
    \end{array}
\end{equation*}
\end{table}

\subsection{Gaussian Relay Networks}
\begin{figure}[t!]
  \begin{center}
  \scalebox{1}{\includegraphics{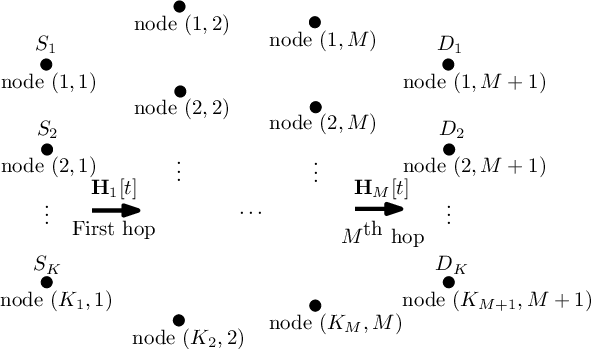}}
  \caption{Layered multi-source multi-hop Gaussian relay networks, where $S_i$ and $D_i$ denote the $i^{\operatorname{th}}$ source and its destination respectively and $K=K_1=K_{M+1}$.}
  \label{fig:Kuser_Mhop}
  \end{center}
\end{figure}

We study a layered Gaussian relay network in Fig. \ref{fig:Kuser_Mhop} consisting of $M+1$ layers with $K_m$ nodes in the $m^{\operatorname{th}}$ layer, where $M\geq 2$.
The nodes in the first layer and the last layer are the sources and the destinations, respectively.
Thus $K=K_1=K_{M+1}$ is the number of S--D pairs.
Let us denote $K_{\min}=\min_{m\in\{1,\cdots,M+1\}}\{K_m\}$ and the $i^{\operatorname{th}}$ node in the $m^{\operatorname{th}}$ layer as  node $(i,m)$, where $i\in\{1,\cdots,K_m\}$ and $m\in\{1,\cdots,M+1\}$.
We assume full-duplex relays so that all relays are able to transmit and receive simultaneously, but the results in this paper can be straightforwardly applied for half-duplex relays by scheduling over hops.

Consider the $m^{\operatorname{th}}$ hop transmission in which the nodes in the $m^{\operatorname{th}}$ layer transmit and the nodes in the $(m+1)^{\operatorname{th}}$ layer receive.
Let $x_{i,m}[t]$ denote the transmit signal of node $(i,m)$ at time $t$ and $y_{j,m}[t]$ denote the received signal of node $(j,m+1)$ at time $t$.
Then the input--output relation of the $m^{\operatorname{th}}$ hop is given by
\begin{equation}
y_{j,m}[t]=\sum_{i=1}^{K_m}h_{ji,m}[t]x_{i,m}[t]+z_{j,m}[t],
\end{equation}
where $h_{ji,m}[t]$ is the complex channel from node $(i,m)$ to node $(j,m+1)$ at time $t$ and $z_{j,m}[t]$ is the additive noise of node $(j,m+1)$ at time $t$.
We assume that $z_{j,m}[t]$'s are independent and identically distributed (i.i.d.) and drawn from $\mathcal{N}_{\mathbb{C}}(0,1)$.
Each node should satisfy the power constraint $P$, i.e., $E(|x_{i,m}[t]|^2)\leq P$.

Let us denote $\mathbf{x}_m[t]=[x_{1,m}[t], \cdots, x_{K_m,m}[t]]^T$ and $\mathbf{y}_m[t]=[y_{1,m}[t], \cdots, y_{K_{m+1},m}[t]]^T$, which are the $K_m\times 1$ dimensional transmit signal vector and the $K_{m+1}\times 1$ dimensional received signal vector of the $m^{\operatorname{th}}$ hop, respectively.
Then the $m^{\operatorname{th}}$ hop transmission can be represented as
\begin{equation}
\mathbf{y}_m[t]=\mathbf{H}_m[t]\mathbf{x}_m[t]+\mathbf{z}_m[t],
\label{eq:input_output_vec}
\end{equation}
where $\mathbf{H}_m[t]$ is the $K_{m+1}\times K_m$ dimensional complex channel matrix of the $m^{\operatorname{th}}$ hop with $[\mathbf{H}_m[t]]_{ji}=h_{ji,m}[t]$ and $\mathbf{z}_m[t]=[z_{1,m}[t],\cdots,z_{K_{m+1},m}[t]]^T$ is the $K_{m+1}\times 1$ dimensional noise vector of the $m^{\operatorname{th}}$ hop.

In this paper, we assume time-varying channels such that $h_{ji,m}[t]$'s are i.i.d. drawn from a continuous probability density function $f_{h}(\cdot)$.
Hence, $f_{\mathbf{H}_m[t]}(\mathbf{H})$ is given by $\prod_{i=1}^{K_m}\prod_{j=1}^{K_{m+1}} f_{h}([\mathbf{H}]_{ji})$, where $f_{\mathbf{H}_m[t]}(\cdot)$ denotes the probability density function of $\mathbf{H}_m[t]$.
We further assume that channel matrices are \emph{isotropically distributed}, i.e., $f_{\mathbf{H}_m[t]}(\mathbf{H})=f_{\mathbf{H}_m[t]}(\mathbf{H}\mathbf{U}_1)=f_{\mathbf{H}_m[t]}(\mathbf{U}_2\mathbf{H})$ for any unitary matrices $\mathbf{U}_1$ and $\mathbf{U}_2$.
We assume that both transmitters and receivers of the $m^{\operatorname{th}}$ hop causally know the global channel state information (CSI) up to the $m^{\operatorname{th}}$ hop.
That is, at time $t_0$, the nodes in the $m^{\operatorname{th}}$ layer know $\{\mathbf{H}_1[t],\cdots,\mathbf{H}_m[t]\}_{t=1}^{t_0}$ if $m\leq M$ and $\{\mathbf{H}_1[t],\cdots,\mathbf{H}_m[t]\}_{t=1}^{t_0}$ if  $m=M+1$.
\begin{remark}
The considered class of channel distributions includes i.i.d. Rayleigh fading in which $h_{ji,m}[t]$ follows $\mathcal{N}_{\mathbb{C}}(0,1)$.
\end{remark}
\begin{remark}
The assumption of time-varying channels can be generalized to block fading with coherence time of $T$ symbols as long as it is big enough such that CSI is available at all relevant nodes. We assume $T=1$ for notational simplicity since our result does not explicitly depend on $T$.
\end{remark}

\subsection{Problem Statement} \label{subsec:prob_state}
Based on the network model, we define a set of length $n$ block codes.
Let $W_i$ be the message of the $i^{\operatorname{th}}$ source uniformly distributed over $\{1,\cdots,2^{nR_i}\}$, where $R_i$ is the rate of the $i^{\operatorname{th}}$ S--D pair.
Then a $(2^{nR_1},\cdots,2^{nR_K};n)$ code consists of the following encoding, relaying, and decoding functions:
\begin{itemize}
\item (Encoding)
For $i\in\{1,\cdots,K\}$, the encoding function of the $i^{\operatorname{th}}$ source, or node $(i,1)$, is given by $f_{i,1,t}:\{1,\cdots,2^{nR_i}\}\to \mathbb{C}$ such that
\begin{equation}
x_{i,1}[t]=f_{i,1,t}(W_i),
\end{equation}
where $t\in\{1,\cdots,n\}$.
\item (Relaying)
For $m\in\{2,\cdots,M\}$ and $i\in\{1,\cdots,K_m\}$, the relaying function of node $(i,m)$ is given by $f_{i,m,t}:\mathbb{C}^{t-1}\to \mathbb{C}$ such that
\begin{equation}
x_{i,m}[t]=f_{i,m,t}\left(y_{i,m-1}[1],\cdots,y_{i,m-1}[t-1]\right),
\end{equation}
where $t\in\{1,\cdots,n\}$.
\item (Decoding)
For $i\in\{1,\cdots,K\}$, the decoding function of the $i^{\operatorname{th}}$ destination, or node $(i,M+1)$, is given by
$g_i:\mathbb{C}^n\to\{1,\cdots,2^{nR_i}\}$ such that
\begin{equation}
\hat{W}_i=g_i\left(y_{i,M}[1],\cdots,y_{i,M}[n]\right).
\end{equation}
\end{itemize}

The probability of error at the $i^{\operatorname{th}}$ destination is given by $P_{e,i}=\Pr(\hat{W}_i\neq W_i)$.
A rate tuple $\left(R_1,\cdots,R_K\right)$ is said to be \emph{achievable} if there exists a sequence of $(2^{nR_1},\cdots,2^{nR_K};n)$ codes with $P_{e,i}\to 0$ as $n\to\infty$ for all $i\in\{1,\cdots,K\}$.
The capacity region $\mathcal{C}$ is the closure of the set of all achievable rate tuples.
In the same manner as for the $K$-user interference channel \cite{Viveck1:08}, we define the DoF region as
\begin{eqnarray}
\!\!\!\!\!\!\!\!\!\!&&\mathcal{D}=\Bigg\{(d_1,\cdots,d_K)\in \mathbb{R}_{+}^K\Bigg|\forall (w_1,\cdots,w_K)\in \mathbb{R}_{+}^K,\nonumber\\
\!\!\!\!\!\!\!\!\!\!&&\sum_{i=1}^Kw_id_i\leq \underset{P\to \infty}{\lim\sup}\left(\sup_{(R_1,\cdots,R_K)\in\mathcal{C}}\sum_{i=1}^K w_i\frac{R_i}{\log P}\right)\Bigg\},
\label{eq:dof_definition}
\end{eqnarray}
where $d_i$ is the DoF of the $i^{\operatorname{th}}$ S--D pair.

\subsection{Multi-antenna and General Message Set}
We also study a more general case in which each node is equipped with multiple antennas and each source has the messages of all destinations.
Let $L_{i,m}$ denote the number of antennas of node $(i,m)$ and $\mathcal{W}_g=\{W_{11},\cdots,W_{K_{M+1}K_1}\}$ denote the set of all $K_1K_{M+1}$ messages, where $W_{ji}$ is the message from the $i^{\operatorname{th}}$ source to the $j^{\operatorname{th}}$ destination and $K_1\neq K_{M+1}$ in general.
Let us denote $L_m=\sum_{i=1}^{K_m} L_{i,m}$ and $L_{\min}=\min_{m\in\{1,\cdots,M+1\}}\big\{L_m\big\}$.
Similar to Section \ref{subsec:prob_state}, the capacity region $\mathcal{C}(\mathcal{W}_g)$ can be defined.
The DoF region is defined as
\begin{eqnarray}
\mathcal{D}(\mathcal{W}_g)\!\!\!\!\!\!\!\!\!&&=\Bigg\{\{d_{ji}\}_{W_{ji}\in\mathcal{W}_g}\in \mathbb{R}_{+}^{K_1K_{M+1}}\Bigg|\forall \{w_{ji}\}_{W_{ji}\in\mathcal{W}_g}\in \mathbb{R}_{+}^{K_1K_{M+1}},\nonumber\\
&&{~~~~~~}\sum_{W_{ji}\in\mathcal{W}_g}w_{ji}d_{ji}\leq \underset{P\to \infty}{\lim\sup}\left(\sup_{\{R_{ji}\}_{W_{ji}\in\mathcal{W}_g}\in\mathcal{C}(\mathcal{W}_g)}\sum_{W_{ji}\in\mathcal{W}_g}w_{ji}\frac{R_{ji}}{\log P}\right)\Bigg\},
\label{eq:dof_def2}
\end{eqnarray}
which is a simple extension of (\ref{eq:dof_definition}).
Here, $d_{ji}$ is the DoF from the $i^{\operatorname{th}}$ source to the $j^{\operatorname{th}}$ destination.

\section{Main Results} \label{sec:result}
Throughout the paper, we study the DoF region of the Gaussian relay network.
We simply state the main results here and derive them in the remainder of the paper.

\begin{theorem} \label{th:dof}
Consider the Gaussian relay network.
Let $K_{\Sigma}$ denote the maximum integer such that $K_{\Sigma}\leq K_{\min}$ and $M/K_{\Sigma}$ is an integer.
Then the set of all $(d_1,\cdots,d_K)$ satisfying
\begin{eqnarray}
&&d_i\leq 1 \mbox{ for }i\in\{1,\cdots,K\},\label{eq:dof1}\\
&&\sum_{i=1}^K d_i\leq K_{\Sigma}\label{eq:dof2}
\end{eqnarray}
is achievable.
\end{theorem}
\begin{proof}
We refer to Section \ref{subsec:dof} for the proof.
\end{proof}

\begin{corollary} \label{co:dof}
Consider the Gaussian relay network.
If $M/K_{\min}$ is an integer, then $\mathcal{D}$ coincides with the DoF region in Theorem \ref{th:dof}, where $K_{\Sigma}=K_{\min}$.
\end{corollary}

\begin{figure}[t!]
  \begin{center}
  \scalebox{1}{\includegraphics{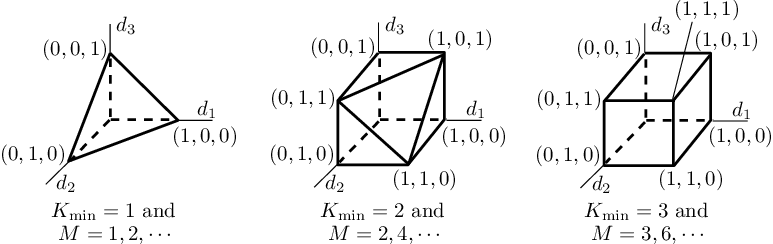}}
  \caption{DoF region $\mathcal{D}$ for the $3$-user Gaussian relay network.}
  \label{fig:dof_region}
  \end{center}
\end{figure}

Notice that Corollary \ref{co:dof} is the first result characterizing the optimal DoF region of multi-source multi-hop networks in which $M/K_{\min}$ is an integer.
The DoF region $\mathcal{D}$ in Corollary \ref{co:dof} coincides with the DoF region assuming perfect cooperation between the relays in each layer and, thus, there is no penalty in DoF due to distributed relays.
This property can be used to characterize the DoF region of more general networks having multi-antenna nodes and general message set.
Fig. \ref{fig:dof_region} plots $\mathcal{D}$ of the $3$-user Gaussian relay network in which $M/K_{\min}$ is an integer.
The sum DoF increases as $K_{\min}$ increases and, in the end, each S--D pair can achieve one DoF simultaneously if $K_{\min}=K$.

\begin{theorem} \label{th:dof_general}
Consider the Gaussian relay network with multi-antenna nodes and general message set.
Let $L_{\Sigma}$ denote the maximum integer such that $L_{\Sigma}\leq L_{\min}$ and $M/L_{\Sigma}$ is an integer.
Then the set of all $\{d_{ji}\}_{W_{ji}\in\mathcal{W}_g}$ satisfying
\begin{eqnarray}
&&\sum_{i=1}^{K_1}d_{ji}\leq L_{j,M+1} \mbox{ for }j\in\{1,\cdots,K_{M+1}\},\label{eq:dof_con1}\\
&&\!\!\sum_{j=1}^{K_{M+1}}d_{ji}\leq L_{i,1} \mbox{ for }i\in\{1,\cdots,K_1\},\label{eq:dof_con2}\\
&&\sum_{i=1}^{K_1}\sum_{j=1}^{K_{M+1}} d_{ji}\leq L_{\Sigma}\label{eq:dof_con3}
\end{eqnarray}
is achievable.
\end{theorem}
\begin{proof}
We refer to Section \ref{subsec:dof_general} for the proof.
\end{proof}

\begin{corollary} \label{co:dof_general}
Consider the Gaussian relay network with multi-antenna nodes and general message set.
If $M/L_{\min}$ is an integer, then $\mathcal{D}(\mathcal{W}_g)$ coincides with the DoF region in Theorem \ref{th:dof_general}, where $L_{\Sigma}=L_{\min}$.
\end{corollary}

\begin{figure}[t!]
  \begin{center}
  \scalebox{1}{\includegraphics{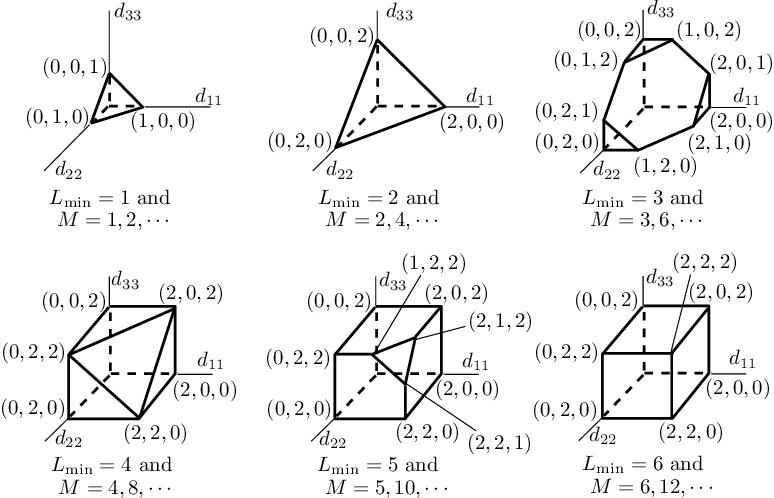}}
  \caption{DoF region $\mathcal{D}(\mathcal{W}_g)$ for the $3$-user Gaussian relay network, where $\mathcal{W}_g=\{W_{11},W_{22},W_{33}\}$ and $L_{i,1}=L_{i,M+1}=2$ for $i\in\{1,2,3\}$.}
  \label{fig:dof_region2}
  \end{center}
\end{figure}

Corollary \ref{co:dof_general} again characterizes $\mathcal{D}(\mathcal{W}_g)$ if $M/L_{\min}$ is an integer, which is the first result showing the optimal DoF region for this class of networks.
The DoF region $\mathcal{D}(\mathcal{W}_g)$ in Corollary \ref{co:dof_general} coincides with DoF region assuming perfect cooperation between the relays in each layer.
Fig. \ref{fig:dof_region2} plots $\mathcal{D}(\mathcal{W}_g)$ of the $3$-user Gaussian relay network in which $\mathcal{W}_g=\{W_{11},W_{22},W_{33}\}$, $L_{i,1}=L_{i,M+1}=2$ for $i\in\{1,2,3\}$, and $M/L_{\min}$ is an integer.

\section{DoF Region for $K$-user $K$-hop Networks} \label{sec:dof_K_K}
To prove the main results, we first study the $K$-user $K$-hop Gaussian relay network in which $K_m=K$ for all $m$.
We propose an AF relaying scheme and derive its achievable DoF region.
This result will be used to show a general achievable DoF region in Section \ref{sec:general_dof}.
In this section, we will be dealing with the case that $K$ is even and refer to Appendix II for odd $K$.

\subsection{Opportunistic Interference Cancellation}
As shown in the introduction, interference-free communication is possible for all S--D pairs if messages are transmitted at time $t_1$ to $t_K$ such that $\prod_{i=1}^K\mathbf{H}_i[t_i]$ becomes a diagonal matrix.
The relays in each layer, however, will have to wait forever in order to group a series of channel matrices perfectly since channel coefficients vary according to a continuous distribution.
To resolve this problem, we first partition the entire channel space of each hop into subsets based on the singular value decomposition (SVD) and then group a series of subsets over $K$ hops.
Before describing our proposed scheme, we define the unordered SVD.

\subsubsection{Unordered SVD}
Let $\mathbf{H}\in\mathbb{C}^{m\times m}$ and $\mathcal{U}_m=\{\mathbf{A}\big|\mathbf{A}\mathbf{A}^{\dagger}=\mathbf{I}_m,\mathbf{A}\in\mathbb{C}^{m\times m}\}$ denote the set of all $m\times m$ dimensional unitary matrices.
First consider the ordered SVD $S_o:\mathbf{H}\to (\mathbf{U}_o,\mathbf{\Sigma}_o,\mathbf{V}_o)$ such that $\mathbf{U}_o\mathbf{\Sigma}_o\mathbf{V}_o^{\dagger}=\mathbf{H}$.
Here, $\mathbf{U}_o$ is the left unitary matrix, $\mathbf{\Sigma}_o$ is the diagonal matrix with ordered singular values from the greatest to the least, and $\mathbf{V}_o$ is the right unitary matrix.\footnote{Singular values are distinct and positive with probability one under the considered class of channel distributions.}
To make the ordered SVD unique, we assume that the first row of $\mathbf{U}_o$ is real and non-negative \cite{Lizhong:02}.

From the ordered SVD, we define the unordered SVD by introducing two random matrices $\mathbf{\Theta}$ and $\mathbf{\Gamma}$.
Define $S:\mathbf{H}\rightarrow (\mathbf{U},\mathbf{\Sigma},\mathbf{V})$ such that
\begin{equation}
S(\mathbf{H})\triangleq(\mathbf{U}_o\mathbf{\Theta}\mathbf{\Gamma},\mathbf{\Gamma}^T\mathbf{\Sigma}_o\mathbf{\Gamma},\mathbf{V}_o\mathbf{\Theta}\mathbf{\Gamma}),
\label{eq:unordered_svd}
\end{equation}
where $(\mathbf{U}_o,\mathbf{\Sigma}_o,\mathbf{V}_o)=S_o(\mathbf{H})$.
Here, $\mathbf{\Gamma}$ is a permutation matrix that is set to be one of $m!$ possible permutations with equal probability and $\mathbf{\Theta}=\operatorname{diag}(e^{j\theta_1},\cdots,e^{j\theta_m})$, where $\theta_i$'s are i.i.d. and uniformly distributed over $[0,2\pi)$.
Hence, for any unitary matrices $\mathbf{U}$, $\mathbf{V}$ and diagonal matrix $\mathbf{\Sigma}$ with $[\mathbf{\Sigma}]_{ii}>0$,  $(\mathbf{U},\mathbf{\Sigma},\mathbf{V})$ can be an instance of $S(\mathbf{H})$ if $\mathbf{U}\mathbf{\Sigma}\mathbf{V}^{\dagger}=\mathbf{H}$.

For a random matrix $\mathbf{H}\in\mathbb{C}^{m\times m}$, let $f_{S(\mathbf{H})}(\mathbf{U},\mathbf{\Sigma},\mathbf{V})$ denote the joint probability density function of $S(\mathbf{H})$.
Since the total number of real dimensions of $\mathcal{U}_m$ is equal to $m^2$ \cite{Lizhong:02}, $f_{S(\mathbf{H})}(\mathbf{U},\mathbf{\Sigma},\mathbf{V})$ is defined over a manifold of $2m^2+m$ real dimensions embedded in $4m^2+m$ dimensional Euclidean space.
Let $f_{\mathbf{U}}(\mathbf{U})$, $f_{\mathbf{\Sigma}}(\mathbf{\Sigma})$, and $f_{\mathbf{V}}(\mathbf{V})$ denote its marginal probability density functions, where $(\mathbf{U},\mathbf{\Sigma},\mathbf{V})=S(\mathbf{H})$.
Then the following lemma holds.

\begin{lemma} \label{lemma:product_pdf}
Suppose that $\mathbf{H}\in\mathbb{C}^{m\times m}$ is isotropically distributed.
Then $f_{S(\mathbf{H})}(\mathbf{U},\mathbf{\Sigma},\mathbf{V})=f_{\mathbf{U}}(\mathbf{U})f_{\mathbf{\Sigma}}(\mathbf{\Sigma})f_{\mathbf{V}}(\mathbf{V})$, where $f_{\mathbf{U}}(\cdot)=f_{\mathbf{V}}(\cdot)=\prod_{i=1}^m\frac{(i-1)!}{2\pi^i}$ and $f_{\mathbf{\Sigma}}(\mathbf{\Sigma})=f_{\mathbf{\Sigma}}(\mathbf{\Gamma}^T\mathbf{\Sigma}\mathbf{\Gamma})$ for any permutation matrix $\mathbf{\Gamma}$.
\end{lemma}
\begin{proof}
We refer to Appendix I for the proof.
\end{proof}

In essence, for isotropically distributed channel matrices, the joint probability density function of $S(\mathbf{H})$ is given by the product of its marginal distributions.
It also shows that $f_{S(\mathbf{H})}(\mathbf{U},\mathbf{\Sigma},\mathbf{V})$'s are the same if their sets of singular values are the same.
This property will be used to show that the probabilities of a series of grouped channel subsets are the same in Lemma \ref{lemma:eq_prob} or asymptotically the same in Appendix II.

\subsubsection{Channel space partitioning}
Let us partition channel spaces of each hop.
Define $\mathcal{B}_{\delta}=\{\delta k\big||k|\leq \alpha,k\in\mathbb{Z}\}$, where $\delta>0$ is the quantization interval and $\alpha\in\mathbb{Z}_+$ is related to the number of quantization points.
Then define $\mathcal{Q}_{\delta}\triangleq\{\mathbf{A}\in\mathbb{C}^{K\times K}\big|\operatorname{real}([\mathbf{A}]_{ij})\in\mathcal{B}_{\delta},\operatorname{imag}([\mathbf{A}]_{ij})\in\mathcal{B}_{\delta}, 1\leq i,j\leq K\}$ and $\mathcal{I}_{\delta}\triangleq\{\mathbf{A}\in\operatorname{diag}(\mathbb{R}^{K\times 1})\big|[\mathbf{A}]_{ii}\in\mathcal{B}_{\delta}, [\mathbf{A}]_{ii}\geq 0, 1\leq i\leq K \}$, where $\operatorname{card}(\mathcal{Q}_{\delta})=(2\alpha+1)^{2K^2}$ and $\operatorname{card}(\mathcal{I}_{\delta})=(\alpha+1)^{K}$.

For $\mathbf{U}_{\delta}\in\mathcal{Q}_{\delta}$, define $\mathcal{Q}(\mathbf{U}_{\delta})\triangleq\{\mathbf{A}\in \mathcal{U}_K\big|-\delta/2\leq \operatorname{real}([\mathbf{A}]_{ij}-[\mathbf{U}_{\delta}]_{ij})<\delta/2,-\delta/2\leq \operatorname{imag}([\mathbf{A}]_{ij}-[\mathbf{U}_{\delta}]_{ij})<\delta/2,1\leq i,j\leq K\}$.
For $\mathbf{\Sigma}_{\delta}\in\mathcal{I}_{\delta}$, define $\mathcal{I}(\mathbf{\Sigma}_{\delta})\triangleq\{\mathbf{A}\in\operatorname{diag}(\mathbb{R}^{K\times 1})\big|-\delta/2\leq [\mathbf{A}]_{ii}-[\mathbf{\Sigma}_{\delta}]_{ii}<\delta/2,1\leq i\leq K\}$.
Then, for $\mathbf{U}_{\delta}\in \mathcal{Q}_{\delta}$, $\mathbf{\Sigma}_{\delta}\in\mathcal{I}_{\delta}$, and $\mathbf{V}_{\delta}\in\mathcal{Q}_{\delta}$, define
\begin{equation}
\mathcal{S}(\mathbf{U}_{\delta},\mathbf{\Sigma}_{\delta},\mathbf{V}_{\delta})\triangleq \mathcal{Q}(\mathbf{U}_{\delta})\times\mathcal{I}(\mathbf{\Sigma}_{\delta})\times\mathcal{Q}(\mathbf{V}_{\delta}).
\label{eq:S_svd}
\end{equation}

The following lemma shows that the probability that $S(\mathbf{H}_m[t])\in\mathcal{S}(\mathbf{U}_{\delta},\mathbf{\Sigma}^{(1)}_{\delta},\mathbf{V}_{\delta})$ is equal to that of $S(\mathbf{H}_m[t])\in\mathcal{S}(\mathbf{V}_{\delta},\mathbf{\Sigma}^{(2)}_{\delta},\mathbf{U}_{\delta})$ if the diagonal elements of $\mathbf{\Sigma}_{\delta}^{(1)}$ is a permutation of those of $\mathbf{\Sigma}_{\delta}^{(2)}$.

\begin{lemma} \label{lemma:eq_prob}
Suppose that $\mathbf{H}\in\mathbb{C}^{m\times m}$ is isotropically distributed.
For $\mathbf{\Sigma}_{\delta}^{(1)}\in\mathcal{I}_{\delta}$ and $\mathbf{\Sigma}_{\delta}^{(2)}\in\mathcal{I}_{\delta}$, if there exists a permutation matrix $\mathbf{\Gamma}$ such that $\mathbf{\Sigma}_{\delta}^{(2)}=\mathbf{\Gamma}^T\mathbf{\Sigma}_{\delta}^{(1)}\mathbf{\Gamma}$,
then $\Pr(S(\mathbf{H}_m[t])\in\mathcal{S}(\mathbf{U}_{\delta},\mathbf{\Sigma}^{(1)}_{\delta},\mathbf{V}_{\delta}))=\Pr(S(\mathbf{H}_m[t])\in\mathcal{S}(\mathbf{V}_{\delta},\mathbf{\Sigma}^{(2)}_{\delta},\mathbf{U}_{\delta}))$
for all $\mathbf{U}_{\delta}\in\mathcal{Q}_{\delta}$, and $\mathbf{V}_{\delta}\in\mathcal{Q}_{\delta}$.
\end{lemma}
\begin{proof}
We have
\begin{eqnarray}
\!\!\!\!\!\!\!\!\!\!\!\!\!\!&&\Pr(S(\mathbf{H}_m[t])\in\mathcal{S}(\mathbf{U}_{\delta},\mathbf{\Sigma}^{(1)}_{\delta},\mathbf{V}_{\delta}))\nonumber\\
\!\!\!\!\!\!\!\!\!\!\!\!\!\!&&\overset{(a)}{=}\int_{\mathbf{U}\in\mathcal{Q}(\mathbf{U}_{\delta})}f_{\mathbf{U}}(\mathbf{U})d\mathbf{U}\int_{\mathbf{\Sigma}\in\mathcal{I}(\mathbf{\Sigma}^{(1)}_{\delta})}f_{\mathbf{\Sigma}}(\mathbf{\Sigma})d\mathbf{\Sigma}\nonumber\\
&&{~~~~}\cdot\int_{\mathbf{V}\in\mathcal{Q}(\mathbf{V}_{\delta})}f_{\mathbf{V}}(\mathbf{V})d\mathbf{V}\nonumber\\
\!\!\!\!\!\!\!\!\!\!\!\!\!\!&&\overset{(b)}{=}\int_{\mathbf{V}\in\mathcal{Q}(\mathbf{V}_{\delta})}f_{\mathbf{V}}(\mathbf{V})d\mathbf{V}\int_{\mathbf{\Sigma}'\in\mathcal{I}(\mathbf{\Sigma}^{(2)}_{\delta})}f_{\mathbf{\Sigma}}(\mathbf{\Sigma}')d\mathbf{\Sigma}'\nonumber\\
&&{~~~~}\cdot\int_{\mathbf{U}\in\mathcal{Q}(\mathbf{U}_{\delta})}f_{\mathbf{U}}(\mathbf{U})d\mathbf{U}\nonumber\\
\!\!\!\!\!\!\!\!\!\!\!\!\!\!&&\overset{(c)}{=}\Pr(S(\mathbf{H}_m[t])\in\mathcal{S}(\mathbf{V}_{\delta},\mathbf{\Sigma}^{(2)}_{\delta},\mathbf{U}_{\delta})),
\end{eqnarray}
where $(a)$ holds from Lemma \ref{lemma:product_pdf},
$(b)$ is obtained by setting $\mathbf{\Sigma}'=\mathbf{\Gamma}^T\mathbf{\Sigma}\mathbf{\Gamma}$ whose Jacobian is one,
and $(c)$ holds since $f_{\mathbf{U}}(\cdot)=f_{\mathbf{V}}(\cdot)$ and $f_{\mathbf{\Sigma}}(\mathbf{\Sigma})=f_{\mathbf{\Sigma}}(\mathbf{\Gamma}^T\mathbf{\Sigma}\mathbf{\Gamma})$, which is the result of Lemma \ref{lemma:product_pdf}.
In conclusion, Lemma \ref{lemma:eq_prob} holds.
\end{proof}

This lemma is crucially important because it will be used to show that the probabilities of grouped channel subsets are the same.
Otherwise, a constant fraction of channel instances remains unused and this may degrade DoF.

\subsubsection{Proposed AF relaying}
First, we divide a block into $B+K-1$ sub-blocks having length $n_B$ for each sub-block, where $n_B=\frac{n}{B+K-1}$.
The relay nodes in each layer will receive length-$n_B$ signals from the previous layer and then amplify and forward them to the next layer with one sub-block delay.
That is, each length-$n_B$ signal transmitted by the sources is received by the destinations with $K-1$ sub-block delay.
Hence the number of effective sub-blocks is equal to $B$ and the overall rate is given by $\frac{B}{B+K-1}R_k$.
As $n\to\infty$, the fractional rate loss $1-\frac{B}{B+K-1}$ will be negligible because we can make both $n_B$ and $B$ large enough.
Thus we omit the sub-block index in describing the proposed scheme.

For $m\in\{1,\cdots,K\}$, define
\begin{eqnarray}
\!\!\!\!\!\!\!\!\!\!\!\!\!\!\!&&\mathcal{T}_{m}(\mathbf{U}_{\delta},\mathbf{\Sigma}_{\delta},\mathbf{V}_{\delta})\nonumber\\
\!\!\!\!\!\!\!\!\!\!\!\!\!\!\!&&\triangleq\big\{t\big|S(\mathbf{H}_m[t])\in\mathcal{S}(\mathbf{U}_{\delta},\mathbf{\Sigma}_{\delta},\mathbf{V}_{\delta}),t\in\{1,\cdots, n_B\}\big\},
\end{eqnarray}
which is the set of time indices of the $m^{\operatorname{th}}$ hop such that $S(\mathbf{H}_m[t])$ is in $\mathcal{S}(\mathbf{U}_{\delta},\mathbf{\Sigma}_{\delta},\mathbf{V}_{\delta})$.
For transmission, each node in the $m^{\operatorname{th}}$ layer will use $N(\mathbf{U}_{\delta},\mathbf{\Sigma}_{\delta},\mathbf{V}_{\delta})$ time indices in $\mathcal{T}_{m}(\mathbf{U}_{\delta},\mathbf{\Sigma}_{\delta},\mathbf{V}_{\delta})$.
The detailed procedure is as follows:
\begin{itemize}
\item (Encoding)
\\
For all $(\mathbf{U}_{\delta},\mathbf{\Sigma}_{\delta},\mathbf{V}_{\delta})\in\mathcal{Q}_{\delta}\times\mathcal{I}_{\delta}\times\mathcal{Q}_{\delta}$, the sources transmit their messages with a standard Gaussian codebook satisfying average power $P$ using $N(\mathbf{U}_{\delta},\mathbf{\Sigma}_{\delta},\mathbf{V}_{\delta})$ time indices in $\mathcal{T}_1(\mathbf{U}_{\delta},\mathbf{\Sigma}_{\delta},\mathbf{V}_{\delta})$.
If $\operatorname{card}(\mathcal{T}_1(\mathbf{U}_{\delta},\mathbf{\Sigma}_{\delta},\mathbf{V}_{\delta}))<N(\mathbf{U}_{\delta},\mathbf{\Sigma}_{\delta},\mathbf{V}_{\delta})$ for any $(\mathbf{U}_{\delta},\mathbf{\Sigma}_{\delta},\mathbf{V}_{\delta})$, it declares an error.
\item (Relaying for $m=\{2,\cdots,K\}$)
\\
For all $(\mathbf{U}_{\delta},\mathbf{\Sigma}_{\delta},\mathbf{V}_{\delta})\in\mathcal{Q}_{\delta}\times\mathcal{I}_{\delta}\times\mathcal{Q}_{\delta}$, the nodes in the $m^{\operatorname{th}}$ layer amplify and forward their received signals that are received during $\mathcal{T}_{m-1}(\mathbf{V}_{\delta},\mathbf{P}^T\mathbf{\Sigma}_{\delta}\mathbf{P},\mathbf{U}_{\delta})$ using $N(\mathbf{U}_{\delta},\mathbf{\Sigma}_{\delta},\mathbf{V}_{\delta})$ time indices in $\mathcal{T}_m(\mathbf{U}_{\delta},\mathbf{\Sigma}_{\delta},\mathbf{V}_{\delta})$, where $\mathbf{P}=[[\mathbf{0}_{1\times(K-1)},1]^T,[\mathbf{I}_{K-1},\mathbf{0}_{(K-1)\times1}]^T]^T$.
Specifically, $\mathbf{x}_m[t_m]=\gamma_m\mathbf{y}_{m-1}[t_{m-1}]$, where $t_m\in\mathcal{T}_m(\mathbf{U}_{\delta},\mathbf{\Sigma}_{\delta},\mathbf{V}_{\delta})$ and $t_{m-1}\in\mathcal{T}_{m-1}(\mathbf{V}_{\delta},\mathbf{P}^T\mathbf{\Sigma}_{\delta}\mathbf{P},\mathbf{U}_{\delta})$.
Here, $\gamma_m>0$ is the amplification factor of the $m^{\operatorname{th}}$ hop that should be set to satisfy the power constraint $P$.
If $\operatorname{card}(\mathcal{T}_m(\mathbf{U}_{\delta},\mathbf{\Sigma}_{\delta},\mathbf{V}_{\delta}))<N(\mathbf{U}_{\delta},\mathbf{\Sigma}_{\delta},\mathbf{V}_{\delta})$ for any $(\mathbf{U}_{\delta},\mathbf{\Sigma}_{\delta},\mathbf{V}_{\delta})$, it declares an error.
\item (Decoding)
\\
The destinations decode their messages from the received signals for all $(\mathbf{U}_{\delta},\mathbf{\Sigma}_{\delta},\mathbf{V}_{\delta})\in\mathcal{Q}_{\delta}\times\mathcal{I}_{\delta}\times\mathcal{Q}_{\delta}$.
\end{itemize}

\begin{remark}
Because of $\mathbf{\Gamma}$ and $\mathbf{\Theta}$ in (\ref{eq:unordered_svd}), $S(\mathbf{H})$ is random.
Hence, in order to know $S(\mathbf{H})$ from $\mathbf{H}$ at relevant nodes, the additional information about $\mathbf{\Gamma}$ and $\mathbf{\Theta}$ should be shared by the nodes.
Note that these information can be shared with marginal overhead for block fading with big enough $T$.
\end{remark}

For the proposed scheme, messages are transmitted through a series of particular time indices $t_1$ to $t_K$ such that
\begin{equation}
S(\mathbf{H}_m[t_m])\in\begin{cases} \mathcal{S}(\mathbf{U}_{\delta},\mathbf{P}^{m-1}\mathbf{\Sigma}_{\delta}(\mathbf{P}^T)^{m-1},\mathbf{V}_{\delta}) & \!\!\!\mbox{ for odd }m,\\
\mathcal{S}(\mathbf{V}_{\delta},\mathbf{P}^{m-1}\mathbf{\Sigma}_{\delta}(\mathbf{P}^T)^{m-1},\mathbf{U}_{\delta}) & \!\!\!\mbox{ for even }m.
\label{eq:paired_channels}
\end{cases}
\end{equation}
Because of the permutation matrix $\mathbf{P}$, the diagonal elements of $\mathbf{P}\mathbf{\Sigma}_{\delta}\mathbf{P}^T$ is cyclic shifted from the diagonal elements of $\mathbf{\Sigma}_{\delta}$.
Hence, interference-free communication is possible as the quantization interval $\delta$ converges to zero, which will be proved in the next subsection.

Let $E_{1,i}$ denote the encoding or relaying error event and $E_{2,i}$ denote the decoding error event of the $i^{\operatorname{th}}$ S--D pair.
Notice that $E_{1,i}$ occurs if $\operatorname{card}(\mathcal{T}_m(\mathbf{U}_{\delta},\mathbf{\Sigma}_{\delta},\mathbf{V}_{\delta}))<N(\mathbf{U}_{\delta},\mathbf{\Sigma}_{\delta},\mathbf{V}_{\delta})$ for any $(\mathbf{U}_{\delta},\mathbf{\Sigma}_{\delta},\mathbf{V}_{\delta})$ or $m$.
From the union bound, $P^{(n_B)}_{e,i}\leq\Pr(E_{1,i})+\Pr(E_{2,i})$.

\subsection{Achievable DoF Region}
In this subsection, we derive the achievable DoF region of the proposed scheme.
We will use the shorthand notation $P(\mathbf{U}_{\delta},\mathbf{\Sigma}_{\delta},\mathbf{V}_{\delta})$ to denote $\Pr(S(\mathbf{H}_m[t])\in\mathcal{S}(\mathbf{U}_{\delta},\mathbf{\Sigma}_{\delta},\mathbf{V}_{\delta}))$, which is valid since $\Pr(S(\mathbf{H}_m[t])\in\mathcal{S}(\mathbf{U}_{\delta},\mathbf{\Sigma}_{\delta},\mathbf{V}_{\delta}))$ is the same for all $m$ and $t$.
We first introduce the following lemma.

\begin{lemma}[Csisz{\'a}r and K{\"o}rner] \label{lemma:large_num}
The probability that
\begin{equation}
\left|\frac{1}{n_B}\operatorname{card}(\mathcal{T}_m(\mathbf{U}_{\delta},\mathbf{\Sigma}_{\delta},\mathbf{V}_{\delta}))-P(\mathbf{U}_{\delta},\mathbf{\Sigma}_{\delta},\mathbf{V}_{\delta})\right|\leq \epsilon
\end{equation}
for all $\mathbf{U}_{\delta}\in\mathcal{Q}_{\delta}$, $\mathbf{\Sigma}_{\delta}\in\mathcal{I}_{\delta}$, and $\mathbf{V}_{\delta}\in\mathcal{Q}_{\delta}$ is greater than $1-\operatorname{card}(\mathcal{Q}_{\delta})^2\operatorname{card}(\mathcal{I}_{\delta})/(4n_B \epsilon^2)$.
\end{lemma}
\begin{proof}
We refer to Lemma 2.12 in \cite{Csiszar:81} for the proof.
\end{proof}

The following theorem shows that each S--D pair can achieve one DoF simultaneously if $M=K=K_m$.
This theorem will be used to prove Theorems \ref{th:dof} and \ref{th:dof_general} in Section \ref{sec:general_dof}.

\begin{theorem} \label{th:dof_K_K}
Consider the Gaussian relay network in which $M=K=K_m$ for all $m$.
Then the set of all $(d_1,\cdots,d_K)$ satisfying
\begin{equation}
d_i\leq 1 \mbox{ for }i\in\{1,\cdots,K\}
\end{equation}
is achievable.
\end{theorem}
\begin{proof}
We will prove the case where $K$ is even and refer to Appendix II for the proof of odd $K$.
From Lemma \ref{lemma:large_num}, we set $N(\mathbf{U}_{\delta},\mathbf{\Sigma}_{\delta},\mathbf{V}_{\delta})=\max\{\lfloor n_B(P(\mathbf{U}_{\delta},\mathbf{\Sigma}_{\delta},\mathbf{V}_{\delta})-\epsilon) \rfloor,0\}$.
Hence
\begin{eqnarray}
\Pr(E_{1,i})\!\!\!\!\!\!\!\!\!&&\leq \frac{K(2\alpha+1)^{4K^2}(\alpha+1)^{K}}{4n_B\epsilon^2}\nonumber\\
&&\leq\frac{K2^K3^{4K^2}\alpha^{5K^2}}{4n_B\epsilon^2},
\end{eqnarray}
where we use $\operatorname{card}(\mathcal{Q}_{\delta})=(2\alpha+1)^{2K^2}$, $\operatorname{card}(\mathcal{I}_{\delta})=(\alpha+1)^K$, and the union bound.
Then $N(\mathbf{U}_{\delta},\mathbf{\Sigma}_{\delta},\mathbf{V}_{\delta})$ is equal to $N(\mathbf{V}_{\delta},\mathbf{P}^T\mathbf{\Sigma}_{\delta}\mathbf{P},\mathbf{U}_{\delta})$ because $P(\mathbf{U}_{\delta},\mathbf{\Sigma}_{\delta},\mathbf{V}_{\delta})=P(\mathbf{V}_{\delta},\mathbf{P}^T\mathbf{\Sigma}_{\delta}\mathbf{P},\mathbf{U}_{\delta})$, which is the result of Lemma \ref{lemma:eq_prob}.
Hence, the nodes in the $m^{\operatorname{th}}$ layer are able to amplify and forward $N(\mathbf{V}_{\delta},\mathbf{P}^T\mathbf{\Sigma}_{\delta}\mathbf{P},\mathbf{U}_{\delta})$ received signals by using the time indices in $\mathcal{T}_m(\mathbf{U}_{\delta},\mathbf{\Sigma}_{\delta},\mathbf{V}_{\delta})$ if $E_{1,i}$ does not occur.

Recall that messages are transmitted through a series of particular time indices $t_1$ to $t_K$ satisfying (\ref{eq:paired_channels}).
Then, by letting
\begin{equation}
\mathbf{H}_{\delta,m}\triangleq\begin{cases} \mathbf{U}_{\delta}\mathbf{P}^{m-1}\mathbf{\Sigma}_{\delta}(\mathbf{P}^T)^{m-1}\mathbf{V}^{\dagger}_{\delta} &\mbox{ for odd } m,\\
\mathbf{V}_{\delta}\mathbf{P}^{m-1}\mathbf{\Sigma}_{\delta}(\mathbf{P}^T)^{m-1}\mathbf{U}^{\dagger}_{\delta} &\mbox{ for even } m,
\end{cases}
\end{equation}
$\mathbf{H}_m[t_m]$ can be represented as $\mathbf{H}_{\delta,m}+\mathbf{\Delta}_m$, where $\mathbf{\Delta}_m$ is the quantization error matrix of $\mathbf{H}_m[t_m]$ with respect to $\mathbf{H}_{\delta,m}$.
Since $\mathbf{x}_m[t_m]=\gamma_m \mathbf{y}_{m-1}[t_{m-1}]$, the received signal vector of the last hop is given by
\begin{equation}
\mathbf{y}_K[t_K]=\Big(\prod_{j=2}^K\gamma_{j}\Big)\Big(\prod_{j=1}^K (\mathbf{H}_{\delta,j}+\mathbf{\Delta}_j)\Big)\mathbf{x}_1[t_1]+\mathbf{z}_{AF}+\mathbf{z}_K[t_K],
\end{equation}
where
\begin{equation}
\mathbf{z}_{AF}=\sum_{j=2}^K\Big(\prod_{k=j}^K\gamma_{k}\Big)\Big(\prod_{k=j}^K (\mathbf{H}_{\delta,k}+\mathbf{\Delta}_k)\Big)\mathbf{z}_{j-1}[t_{j-1}]
\label{eq:n_af}
\end{equation}
denotes the accumulated noise due to AF relaying.
Let
\begin{equation}
\operatorname{SINR}^{\min}_i(\mathbf{U}_{\delta},\mathbf{\Sigma}_{\delta},\mathbf{V}_{\delta})\triangleq\min_{\{\mathbf{\Delta}_m\}_{m=1}^K}\{\operatorname{SINR}_i(y_{i,K}[t_K])\},
\end{equation}
where $\operatorname{SINR}_i(y_{i,K}[t_K])$ is the signal-to-noise-and-interference ratio (SINR) of the $i^{\operatorname{th}}$ destination assuming that $S(\mathbf{H}_1[t_1])\in\mathcal{S}(\mathbf{U}_{\delta},\mathbf{\Sigma}_{\delta},\mathbf{V}_{\delta})$.
Therefore, since each source uses a standard Gaussian codebook, an achievable rate of the $i^{\operatorname{th}}$ S--D pair is lower bounded by
\begin{eqnarray}
R_{i,\delta}\!\!\!\!\!\!\!\!\!\!&&\geq\frac{1}{n_B}\sum_{\overset{(\mathbf{U}_{\delta},\mathbf{\Sigma}_{\delta},\mathbf{V}_{\delta})}{\in\mathcal{Q}_{\delta}\times\mathcal{I}_{\delta}\times\mathcal{Q}_{\delta}}}\!\!\log(1+\operatorname{SINR}^{\min}_i(\mathbf{U}_{\delta},\mathbf{\Sigma}_{\delta},\mathbf{V}_{\delta})) N(\mathbf{U}_{\delta},\mathbf{\Sigma}_{\delta},\mathbf{V}_{\delta})\nonumber\\
&&\overset{(a)}{\geq}\!\!\!\!\sum_{\overset{(\mathbf{U}_{\delta},\mathbf{\Sigma}_{\delta},\mathbf{V}_{\delta})}{\in\mathcal{Q}_{\delta}\times\mathcal{I}_{\delta}\times\mathcal{Q}_{\delta}}}\!\!\!\!\log(1+\operatorname{SINR}^{\min}_i(\mathbf{U}_{\delta},\mathbf{\Sigma}_{\delta},\mathbf{V}_{\delta}))P(\mathbf{U}_{\delta},\mathbf{\Sigma}_{\delta},\mathbf{V}_{\delta})-\epsilon'\!\!\!\!\sum_{\overset{(\mathbf{U}_{\delta},\mathbf{\Sigma}_{\delta},\mathbf{V}_{\delta})}{\in\mathcal{Q}_{\delta}\times\mathcal{I}_{\delta}\times\mathcal{Q}_{\delta}}}\!\!\log(1+\operatorname{SINR}^{\min}_i(\mathbf{U}_{\delta},\mathbf{\Sigma}_{\delta},\mathbf{V}_{\delta}))\nonumber\\
&&\overset{(b)}{\geq}\!\!\!\!\sum_{\overset{(\mathbf{U}_{\delta},\mathbf{\Sigma}_{\delta},\mathbf{V}_{\delta})}{\in\mathcal{Q}_{\delta}\times\mathcal{I}_{\delta}\times\mathcal{Q}_{\delta}}}\!\!\!\!\log(1+\operatorname{SINR}^{\min}_i(\mathbf{U}_{\delta},\mathbf{\Sigma}_{\delta},\mathbf{V}_{\delta}))P(\mathbf{U}_{\delta},\mathbf{\Sigma}_{\delta},\mathbf{V}_{\delta})\nonumber\\
&&{~~~~}-2^K3^{4K^2}\epsilon'\alpha^{5K^2}\!\!\!\!\underset{\overset{(\mathbf{U}_{\delta},\mathbf{\Sigma}_{\delta},\mathbf{V}_{\delta})}{\in\mathcal{Q}_{\delta}\times\mathcal{I}_{\delta}\times\mathcal{Q}_{\delta}}}{\max}\!\!\{\log(1+\operatorname{SINR}^{\min}_i(\mathbf{U}_{\delta},\mathbf{\Sigma}_{\delta},\mathbf{V}_{\delta}))\}
\label{eq:rate_lower}
\end{eqnarray}
with an arbitrarily small probability of decoding error, i.e., $P\big(E_{2,i}\big)\to 0$ as $n_B\to \infty$, where $\epsilon'=\epsilon+1/n_B$.
Here, $(a)$ holds since $N(\mathbf{U}_{\delta},\mathbf{\Sigma}_{\delta},\mathbf{V}_{\delta})\geq n_B(P(\mathbf{U}_{\delta},\mathbf{\Sigma}_{\delta},\mathbf{V}_{\delta})-\epsilon')$ and $(b)$ holds since $\operatorname{card}(\mathcal{Q}_{\delta})\leq (2\alpha+1)^{2K^2}$ and $\operatorname{card}(\mathcal{I}_{\delta})\leq (\alpha+1)^{K}$.
Let $\operatorname{SINR}_i(\mathbf{U}_{\delta},\mathbf{\Sigma}_{\delta},\mathbf{V}_{\delta})$ be the SINR of the $i^{\operatorname{th}}$ destination assuming $\mathbf{H}_m[t_m]=\mathbf{H}_{\delta,m}$, which is a function of $\mathbf{U}_{\delta}$, $\mathbf{\Sigma}_{\delta}$, and $\mathbf{V}_{\delta}$.
Then,
\setcounter{equation}{28}
\begin{eqnarray}
&&\max_{\overset{(\mathbf{U}_{\delta},\mathbf{\Sigma}_{\delta},\mathbf{V}_{\delta})}{\in\mathcal{Q}_{\delta}\times\mathcal{I}_{\delta}\times\mathcal{Q}_{\delta}}}\{\log(1+\operatorname{SINR}^{\min}_i(\mathbf{U}_{\delta},\mathbf{\Sigma}_{\delta},\mathbf{V}_{\delta}))\}\nonumber\\
&&\leq \max_{\overset{(\mathbf{U}_{\delta},\mathbf{\Sigma}_{\delta},\mathbf{V}_{\delta})}{\in\mathcal{Q}_{\delta}\times\mathcal{I}_{\delta}\times\mathcal{Q}_{\delta}}}\{\log(1+\operatorname{SINR}_i(\mathbf{U}_{\delta},\mathbf{\Sigma}_{\delta},\mathbf{V}_{\delta}))\}\nonumber\\
&&\leq\max_{\mathbf{\Sigma}_{\delta}\in\mathcal{I}_{\delta}}\left\{\log\left(1+\big(\prod_{j=1}^{K}[\mathbf{\Sigma}_{\delta}]_{jj}\big)^2P\right)\right\}\nonumber\\
&&\leq\log(1+(\delta\alpha)^{2K}P),
\label{eq:rate_upper}
\end{eqnarray}
where the first inequalilty holds since $\operatorname{SINR}^{\min}_i(\mathbf{U}_{\delta},\mathbf{\Sigma}_{\delta},\mathbf{V}_{\delta})$ is less than or equal to $\operatorname{SINR}_i(\mathbf{U}_{\delta},\mathbf{\Sigma}_{\delta},\mathbf{V}_{\delta})$, the second inequality holds from $\prod_{j=1}^K\mathbf{H}_{\delta,j}=\left(\prod_{j=1}^K[\mathbf{\Sigma}_{\delta}]_{jj}\right)\mathbf{I}_K$ and assuming $\mathbf{z}_{AF}=\mathbf{0}_{K\times 1}$ gives an upper bound on the achevable rate, and the third inequality holds since $|[\mathbf{\Sigma}_{\delta}]_{jj}|\leq \delta \alpha$.

Now we set $\delta=n_B^{-1/(32K^2)}$, $\alpha=n_B^{1/(16K^2)}$, and $\epsilon=n_B^{-1/3}$, which are functions of $n_B$. Then
\begin{eqnarray}
&&\delta=n_B^{-1/(32K^2)}\to 0
\label{eq:condition1}\\
&&2^K3^{4K^2}\epsilon'\alpha^{5K^2}\log(1+(\delta\alpha)^{2K}P)\nonumber\\
&&=2^K3^{4K^2}(n_B^{-1/48}+n_B^{-11/16})\nonumber\\
&&{~~}\cdot\log(1+Pn_B^{1/(16K)})\to 0
\label{eq:condition2}\\
&&\delta \alpha=n_B^{1/(32K^2)}\to \infty
\label{eq:condition3}\\
&&\frac{K2^K3^{4K^2}\alpha^{5K^2}}{4n_B\epsilon^2}=\frac{K2^K3^{4K^2}}{4}n_B^{-1/48}\to 0
\label{eq:condition4}
\end{eqnarray}
as $n_B\to\infty$.
The first condition guarantees an arbitrarily small quantization error, the second condition guarantees an arbitrarily small rate loss due to the randomness of channel realizations, the third condition is needed to use almost all channel instances for transmission, and the fourth condition guarantees an arbitrarily small probability of encoding and relaying error.

Since we separately quantize the left unitary matrix, the singular value matrix, and the right unitary matrix, from (\ref{eq:paired_channels}), $S(\mathbf{H}_m[t_m])$ converges to $(\mathbf{U}_{\delta},\mathbf{P}^{m-1}\mathbf{\Sigma}_{\delta}(\mathbf{P}^T)^{m-1},\mathbf{V}_{\delta})$ for odd $m$ and $(\mathbf{V}_{\delta},\mathbf{P}^{m-1}\mathbf{\Sigma}_{\delta}(\mathbf{P}^T)^{m-1},\mathbf{U}_{\delta})$ for even $m$ as $\delta\to 0$.
Hence $\mathbf{H}_m[t_m]$ converges to $\mathbf{H}_{\delta,m}$, equivalently $\mathbf{\Delta}_m$ converges to the all-zero matrix as $\delta\to0$.
Therefore,
\begin{eqnarray}
&&\lim_{\delta\to0}\operatorname{SINR}^{\min}_i(\mathbf{U}_{\delta},\mathbf{\Sigma}_{\delta},\mathbf{V}_{\delta})\nonumber\\
&&\geq\frac{(\prod_{j=2}^K\gamma^2_{j})\big(\prod_{j=1}^K[\mathbf{\Sigma}]_{jj}\big)^2P}{1+\sum_{j=2}^K\big(\prod_{k=j}^K\gamma^2_{k}\big)\operatorname{tr}(\mathbf{\Sigma}^2)^{K-j+1}},
\end{eqnarray}
where $\mathbf{\Sigma}$ denotes the singular value matrix of $\mathbf{H}_1[t_1]$.
Here, we use $\lim_{\delta\to 0}\big(\prod_{j=1}^K\mathbf{H}_j[t_j]\big)=\big(\prod_{j=1}^K[\mathbf{\Sigma}]_{jj}\big)\mathbf{I}_K$ and
\begin{eqnarray}
&&\lim_{\delta\to 0}E\left(\big|[\mathbf{z}_{AF}]_i\big|^2\right)\nonumber\\
&&=E\left(\bigg|\sum_{j=2}^K\Big(\prod_{k=j}^K \gamma_k\Big)\Big[\prod_{k=j}^K\mathbf{H}_k[t_k]\Big]_i\mathbf{z}_{j-1}[t_{j-1}]\bigg|^2\right)\nonumber\\
&&=\sum_{j=2}^K\Big(\prod_{k=j}^K\gamma^2_{k}\Big)\bigg\|\Big[\prod_{k=j}^K \mathbf{H}_k[t_k]\Big]_i\bigg\|^2\nonumber\\
&&\leq\sum_{j=2}^K\Big(\prod_{k=j}^K\gamma^2_{k}\Big)\bigg\|\prod_{k=j}^K \mathbf{H}_k[t_k]\bigg\|_F^2\nonumber\\
&&\leq\sum_{j=2}^K\Big(\prod_{k=j}^K\gamma^2_{k}\Big)\prod_{k=j}^K \|\mathbf{H}_k[t_k]\|^2_F\nonumber\\
&&=\sum_{j=2}^K\Big(\prod_{k=j}^K\gamma^2_{k}\Big)\operatorname{tr}(\mathbf{\Sigma}^2)^{K-j+1},
\label{eq:upper_noise_var}
\end{eqnarray}
where the first and second inequalities hold from $\|[\mathbf{A}]_i\|\leq \|\mathbf{A}\|_F$ and $\|\mathbf{A}\mathbf{B}\|_F\leq \|\mathbf{A}\|_F\|\mathbf{B}\|_F$, respectively.
Finally, we have
\begin{eqnarray}
R_i\!\!\!\!\!\!\!\!\!&&=\lim_{n_B\to \infty}R_{i,\delta}\geq\int_{(\mathbf{U},\mathbf{\Sigma},\mathbf{V})}\log\left(1+\frac{(\prod_{j=2}^K\gamma^2_{j})\big(\prod_{j=1}^K[\mathbf{\Sigma}]_{jj}\big)^2P}{1+\sum_{j=2}^K\big(\prod_{k=j}^K\gamma^2_{k}\big)\operatorname{tr}(\mathbf{\Sigma}^2)^{K-j+1}}\right) f_{\mathbf{U}}(\mathbf{U})f_{\mathbf{\Sigma}}(\mathbf{\Sigma})f_{\mathbf{V}}(\mathbf{V})d\mathbf{U}d\mathbf{\Sigma}d\mathbf{V}\nonumber\\
&&= E_{\mathbf{\Sigma}}\left(\log\left(1+\frac{(\prod_{j=2}^K\gamma^2_{j})\big(\prod_{j=1}^K[\mathbf{\Sigma}]_{jj}\big)^2P}{1+\sum_{j=2}^K\big(\prod_{k=j}^K\gamma^2_{k}\big)\operatorname{tr}(\mathbf{\Sigma}^2)^{K-j+1}}\right)\right)
\label{eq:final_rate}
\end{eqnarray}
 is achievable with probability one.

Now consider an achievable DoF region.
For any $c_l>0$ and $c_u>0$,
\begin{equation}
c_l(\log P)^{-1}\leq \|\mathbf{H}_{m}[t_{m}]\|_F^2\leq c_u \log P
\label{eq:channel_P}
\end{equation}
with probability one as $P\to\infty$.
To satisfy the power constraint $P$, we set $\gamma^2_m=(\log P)^{-1}$  for $m\in\{2,\cdots,M\}$.
Then
\begin{eqnarray}
&&\!\!\!\!\!\!\!\!\!\!\!\!\!E(|x_{i,m}[t_m]|^2)\nonumber\\
&&\!\!\!\!\!\!\!\!\!\!\!\!\!=(\log P)^{-1}E(|y_{i,m-1}[t_{m-1}]|^2)\nonumber\\
&&\!\!\!\!\!\!\!\!\!\!\!\!\!=(\log P)^{-1}E(\left|[\mathbf{H}_{m-1}[t_{m-1}]]_i\mathbf{x}_{m-1}[t_{m-1}]+z_{i,m-1}[t_{m-1}]\right|^2)\nonumber\\
&&\!\!\!\!\!\!\!\!\!\!\!\!\!\leq(\log P)^{-1}(\|\mathbf{H}_{m-1}[t_{m-1}]\|_F^2E(\|\mathbf{x}_{m-1}[t_{m-1}]\|^2)+1).\nonumber\\
\end{eqnarray}
Consider the case where $m=2$. We have $E(|x_{i,2}[t_2]|^2)\leq (\log P)^{-1}(\|\mathbf{H}_{1}[t_{1}]\|_F^2KP+1)$ since $E(\|\mathbf{x}_1[t_1]\|^2_F)\leq KP$.
Hence, from (\ref{eq:channel_P}), $E(|x_{i,2}[t_2]|^2)\leq P$ with probability one as $P\to\infty$.
By applying the same analysis recursively, we can show that $E(|x_{i,m}[t_m]|^2)\leq P$ for all $m$ with probability one as $P\to\infty$.
Therefore, from (\ref{eq:final_rate}) and $\gamma^2_m=(\log P)^{-1}$ and by using the facts that $\prod_{j=1}^K[\mathbf{\Sigma}]_{jj}>0$ and $\operatorname{tr}(\mathbf{\Sigma}^2)=\|\mathbf{H}_{1}[t_{1}]\|_F^2\leq c_u \log P$ with probability one, $d_i=\lim_{P\to\infty}R_i/\log P=1$ is achievable with probability one for all $i\in\{1,\cdots,K\}$, which completes the proof.
\end{proof}

\section{DoF Region for General Networks} \label{sec:general_dof}
Based on the result in Section \ref{sec:dof_K_K}, we prove Theorems \ref{th:dof} and \ref{th:dof_general} and Corollaries \ref{co:dof} and \ref{co:dof_general}.

\subsection{DoF Region of Gaussian Relay Networks} \label{subsec:dof}
In this subsection, we prove Theorem \ref{th:dof} and Corollary \ref{co:dof}.
First, consider the DoF region in Theorem \ref{th:dof}.
The DoF region given by (\ref{eq:dof1}) and (\ref{eq:dof2}) has corner points $(d^*_1,\cdots,d^*_K)$ such that $\sum_{i=1}^K d^*_i=K_{\Sigma}$ and $d^*_i\in\{0,1\}$ for all $i\in\{1,\cdots,K\}$.
Hence, to achieve $(d^*_1,\cdots,d^*_K)$, only $K_{\Sigma}$ S--D pairs with $d^*_i=1$ participate in communication.
We can also choose $K_{\Sigma}$ nodes in each of the remaining layers because $K_{\Sigma}\leq K_{\min}\leq K_m$.
As a result, the reduced network consists of $K_{\Sigma}$ nodes in each layer.
Then, we can apply the proposed scheme to this reduced network over $M/K_{\Sigma}$ times because $M/K_{\Sigma}$ is an integer.
Hence one DoF is achievable for each of the corresponding $K_{\Sigma}$ S--D pairs, where we use the result of Theorem \ref{th:dof_K_K}.
Therefore, $(d^*_1,\cdots,d^*_K)$ is achievable.
Note that any point on the dominant face can be achieved by time sharing between corner points.
In conclusion, Theorem \ref{th:dof} holds.

Now consider Corollary \ref{co:dof}.
From the condition that $M/K_{\min}$ is an integer, we have $K_{\Sigma}=K_{\min}$.
Hence, the achievability is straightforward from Theorem \ref{th:dof}.
The converse can be shown from a simple cut-set outer bound.
Let us first consider the cut dividing the $i^{\operatorname{th}}$ source and the rest of nodes.
Then the rate of the $i^{\operatorname{th}}$ S--D pair is upper bounded by $K\times 1$ single-input multiple-output (SIMO) capacity, which gives $d_i\leq 1 \mbox{ for }i\in\{1,\cdots,K\}$.
From the cut dividing the nodes up to the $m^{\operatorname{th}}$ layer and the rest of nodes, $\sum_{i=1}^K R_i$ is upper bounded by $K_{m+1}\times K_m$ multiple-input multiple-output (MIMO) capacity.
Hence we obtain $\sum_{i=1}^K d_i\leq\min\{K_m,K_{m+1}\}$ and considering all $m\in\{1,\cdots,M\}$ gives $\sum_{i=1}^K d_i\leq K_{\min}$.
In conclusion, Corollary \ref{co:dof} holds.

\subsection{Multi-antenna and General Message Set} \label{subsec:dof_general}
In this subsection, we prove Theorem \ref{th:dof_general} and Corollary \ref{co:dof_general}.
First, consider the DoF region in Theorem \ref{th:dof_general}.
Assume a specific order of $K_1K_{M+1}$ messages in $\mathcal{W}_g$.
We can sequentially allocate $\{d_{ji}\}_{W_{ji}\in\mathcal{W}_g}$ according to this order and, for a given $d_{ji}$, we can maximally allocate available DoF to $d_{ji}$ while satisfying (\ref{eq:dof_con1}) to (\ref{eq:dof_con3}).
Then the resulting $\{d^*_{ji}\}_{W_{ji}\in\mathcal{W}_g}$ is one of the corner points of $\mathcal{D}(\mathcal{W}_g)$.
Since each $d^*_{ji}$ is an integer, we can choose $d^*_{ji}$ antennas at the $i^{\operatorname{th}}$ source and $d^*_{ji}$ antennas at the $j^{\operatorname{th}}$ destination and pair them as $d^*_{ji}$ virtual S--D pairs.
As a result, we can establish a total of $\sum_{i=1}^{K_1}\sum_{j=1}^{K_{M+1}} d^*_{ji}$ virtual S--D pairs because $\{d^*_{ji}\}_{W_{ji}\in\mathcal{W}_g}$ satisfies (\ref{eq:dof_con1}) and (\ref{eq:dof_con2}).
We can also choose a total of $\sum_{i=1}^{K_1}\sum_{j=1}^{K_{M+1}} d^*_{ji}$ antennas in each of the remaining layers because $\{d^*_{ji}\}_{W_{ji}\in\mathcal{W}_g}$ satisfies (\ref{eq:dof_con3}) and $L_{\Sigma}\leq L_{\min}\leq L_m$.
The resulting reduced network consists of $\sum_{i=1}^{K_1}\sum_{j=1}^{K_{M+1}} d^*_{ji}$ virtual S--D pairs with $\sum_{i=1}^{K_1}\sum_{j=1}^{K_{M+1}} d^*_{ji}$ relays in each layer.
Then we can apply the proposed scheme to this reduced network over $M/L_{\Sigma}$ times because $M/L_{\Sigma}$ is an integer.
As a result, all virtual S--D pairs can achieve one DoF from the result of Theorem {\ref{th:dof}}, meaning that $\{d^*_{ji}\}_{W_{ji}\in\mathcal{W}_g}$ is achievable.
Note that any point in the dominant face can be achieved by time sharing between corner points, which completes the proof.

Consider Corollary \ref{co:dof_general}.
Because $L_{\Sigma}=L_{\min}$, the achievability is straightforward from Theorem \ref{th:dof_general}.
The converse can be shown from the cut-set outer bound.
From the cut dividing the $j^{\operatorname{th}}$ destination and the rest of nodes, $\sum_{i=1}^{K_1}R_{ji}$ is upper bounded by $L_M\times L_{j,M+1}$ MIMO capacity, which gives $\sum_{i=1}^{K_1}d_{ji}\leq L_{j,M+1} \mbox{ for }j\in\{1,\cdots,K_{M+1}\}$.
From the cut dividing the $i^{\operatorname{th}}$ source and the rest of nodes, $\sum_{j=1}^{K_{M+1}}R_{ji}$ is upper bounded by $L_{i,1}\times L_2$ MIMO capacity, which gives $\sum_{j=1}^{K_{M+1}}d_{ji}\leq L_{i,1} \mbox{ for }i\in\{1,\cdots,K_1\}$.
Lastly, from the cut dividing the nodes up to the $m^{\operatorname{th}}$ layer and the rest of nodes, $\sum_{i=1}^{K_1}\sum_{j=1}^{K_{M+1}} R_{ji}$ is upper bounded by $L_m\times L_{m+1}$ MIMO capacity, which gives $\sum_{i=1}^{K_1}\sum_{j=1}^{K_{M+1}} d_{ji}\leq \min\{L_m,L_{m+1}\}$ for $m\in\{1,\cdots,M\}$.
Hence, we have $\sum_{i=1}^{K_1}\sum_{j=1}^{K_{M+1}} d_{ji}\leq L_{\min}$.
In conclusion, Corollary \ref{co:dof_general} holds.

\section{Concluding Remarks} \label{sec:conclusion}

\subsection{Summary}
In this paper, we study layered $K$-user $M$-hop Gaussian relay networks.
The proposed AF relaying exploits channel fluctuation to cancel the inter-user interference and works for any isotropically distributed channel matrices including i.i.d. Rayleigh fading.
Under this class of channel distributions, we show a general achievable DoF region, which characterizes the optimal DoF region if $M/K_{\min}$ is an integer.
We further consider the DoF region of more general networks with multi-antenna nodes and general message set.
Our achievable DoF region again characterizes the optimal DoF region if $M/L_{\min}$ is an integer.

\subsection{Discussions}
The proposed channel matching using the unordered SVD works basically for i.i.d. channel coefficients whose channel matrices are isotropically distributed.
Specifically, if we take any $K_S\times K_S$ sub-channels at each hop, the probability density functions of these sub-channels should be the same and isotropically distributed, where $K_S\leq K_{\min}$.
When channel coefficients are arbitrarily correlated, these conditions generally do not hold.
However, we can still apply the opportunistic interference cancellation.
For this case, other channel matching may provide a larger achievable DoF region than the proposed matching can depending on channel correlations.

In this paper, we consider opportunistic interference cancellation based on the AF relaying.
Although the proposed scheme achieves the optimal DoF region for a class of networks, if the number of S--D pairs is relatively greater than the number of hops, then applying the interference alignment in \cite{Viveck2:09} at each hop based on the decode-and-forward relaying can provide a larger total DoF than the proposed scheme.
Furthermore, compress-and-forward in \cite{AvestimehrDiggaviTse:08,Lim:00} or compute-and-forward in \cite{Nazer2:09} may also outperform the proposed scheme in finite SNR regime.

\section*{Appendix I\\Probability Density Functions of Unordered SVD}
In this appendix, we prove Lemma \ref{lemma:product_pdf}.
Let $f_{S_o(\mathbf{H})}(\mathbf{U}_o,\mathbf{\Sigma}_o,\mathbf{V}_o)$ denote the joint probability density function of $S_o(\mathbf{H})$.
Since the first row of $\mathbf{U}_o$ is real and non-negative, $f_{S_o(\mathbf{H})}(\mathbf{U}_o,\mathbf{\Sigma}_o,\mathbf{V}_o)$ is defined over $2m^2$ real dimensions \cite{Lizhong:02}.
Consider any unitary matrices $\mathbf{U}^{(1)}$, $\mathbf{U}^{(2)}$, $\mathbf{V}^{(1)}$, $\mathbf{V}^{(2)}$ and any diagonal matrices $\mathbf{\Sigma}^{(1)}$, $\mathbf{\Sigma}^{(2)}$ with distinct and positive diagonal elements such that $\mathbf{\Sigma}^{(2)}=\mathbf{\Gamma}^T\mathbf{\Sigma}^{(1)}\mathbf{\Gamma}$ for a permutation matrix $\mathbf{\Gamma}$.
We have
\begin{eqnarray}
&&f_{S(\mathbf{H})}(\mathbf{U}^{(1)},\mathbf{\Sigma}^{(1)},\mathbf{V}^{(1)})\nonumber\\
&&\overset{(a)}{=}\frac{1}{(2\pi)^mm!}f_{S_o(\mathbf{H})}(\mathbf{U}_o^{(1)},\mathbf{\Sigma}_o^{(1)},\mathbf{V}_o^{(1)})\nonumber\\
&&\overset{(b)}{=}\frac{1}{(2\pi)^mm!J(\mathbf{\Sigma}^{(1)})}f_{\mathbf{H}}(\mathbf{U}^{(1)}\mathbf{\Sigma}^{(1)}\mathbf{V}^{(1)\dagger})\nonumber\\
&&\overset{(c)}{=}\frac{1}{(2\pi)^mm!J(\mathbf{\Sigma}^{(1)})}f_{\mathbf{H}}(\mathbf{U}_1\mathbf{U}^{(1)}\mathbf{\Sigma}^{(1)}\mathbf{V}^{(1)\dagger}\mathbf{U}_2)\nonumber\\
&&\overset{(d)}{=}\frac{1}{(2\pi)^mm!J(\mathbf{\Sigma}^{(2)})}f_{\mathbf{H}}(\mathbf{U}^{(2)}\mathbf{\Sigma}^{(2)}\mathbf{V}^{(2)\dagger})\nonumber\\
&&\overset{(e)}{=}f_{S(\mathbf{H})}(\mathbf{U}^{(2)},\mathbf{\Sigma}^{(2)},\mathbf{V}^{(2)}),
\label{eq:equal_pdf}
\end{eqnarray}
where $(\mathbf{U}_o^{(1)},\mathbf{\Sigma}_o^{(1)},\mathbf{V}_o^{(1)})=S_o(\mathbf{U}^{(1)}\mathbf{\Sigma}^{(1)}\mathbf{V}^{(1)\dagger})$ and
\begin{equation}
J(\mathbf{\Sigma})=\frac{1}{\prod_{i<j}(\lambda^2_i-\lambda_j^2)^2\prod_{i=1}^m\lambda_i}
\end{equation}
denotes the Jacobian from $\mathbf{H}$ to $S_o{(\mathbf{H})}$ \cite{Lizhong:02} and $\lambda_i$ is the $i^{\operatorname{th}}$ largest singular value in $\mathbf{\Sigma}$.
Here, $(a)$ holds since the probability density function of $\mathbf{\Theta}$ in (\ref{eq:unordered_svd}) is given by $f_{\mathbf{\Theta}}(\mathbf{\Theta})=\frac{1}{(2\pi)^m}$, $\mathbf{\Gamma}$ is set to one of the $m!$ candidates, and the Jacobian from $S(\mathbf{H})$ to $S_o(\mathbf{H})$ is one,
$(b)$ is obtained by $\mathbf{U}_o^{(1)}\mathbf{\Sigma}_o^{(1)}\mathbf{V}_o^{(1)\dagger}=\mathbf{U}^{(1)}\mathbf{\Sigma}^{(1)}\mathbf{V}^{(1)\dagger}$,
$(c)$ holds for any unitary matrices $\mathbf{U}_1$ and $\mathbf{U}_2$,
$(d)$ holds by setting $\mathbf{U}_1=\mathbf{U}^{(2)}\mathbf{\Gamma}^T\mathbf{U}^{(1)\dagger}$ and $\mathbf{U}_2=\mathbf{V}^{(1)}\mathbf{\Gamma}\mathbf{V}^{(2)\dagger}$ and from the fact that $J(\mathbf{\Sigma}^{(1)})=J(\mathbf{\Sigma}^{(2)})$,
$(e)$ holds by the same steps showing that $f_{S(\mathbf{H})}(\mathbf{U}^{(1)},\mathbf{\Sigma}^{(1)},\mathbf{V}^{(1)})=\frac{1}{2m!J(\mathbf{\Sigma}^{(1)})}f_{\mathbf{H}}(\mathbf{U}^{(1)}\mathbf{\Sigma}^{(1)}\mathbf{V}^{(1)\dagger})$.
From (\ref{eq:equal_pdf}), $f_{S(\mathbf{H})}(\mathbf{U},\mathbf{\Sigma},\mathbf{V})$ can be represented as $f_{\mathbf{U}}(\mathbf{U})f_{\mathbf{\Sigma}}(\mathbf{\Sigma})f_{\mathbf{V}}(\mathbf{V})$, where $f_{\mathbf{\Sigma}}(\mathbf{\Sigma})$ is given by $f_{\mathbf{\Sigma}}(\mathbf{\Gamma}^T\mathbf{\Sigma}\mathbf{\Gamma})$ and $f_{\mathbf{U}}(\cdot)=f_{\mathbf{V}}(\cdot)=\prod_{i=1}^m\frac{(i-1)!}{2\pi^i}$ since the volume of $\mathcal{U}_m$ is given by $\prod_{i=1}^m\frac{2\pi^i}{(i-1)!}$ \cite{Lizhong:02}.
In conclusion, Lemma \ref{lemma:product_pdf} holds.

\section*{Appendix II\\DoF Region for $K$-user $K$-hop Networks in which $K$ Is Odd}
\begin{figure}[t!]
  \begin{center}
  \scalebox{1.1}{\includegraphics{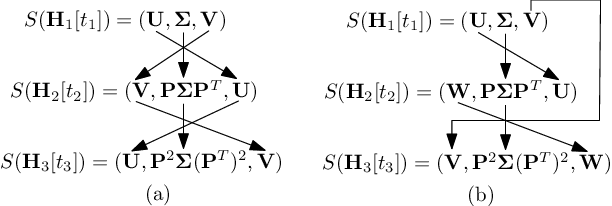}}
  \caption{Channel grouping rules used for even $K$ (a) and odd $K$ (b), where the quantization effect is ignored for simple explanation.}
  \label{fig:odd_K}
  \end{center}
\end{figure}

In this appendix, we prove that Theorem \ref{th:dof_K_K} holds for odd $K$, where $K\geq 3$.
The DoF region in Theorem \ref{th:dof_K_K} is trivially achievable if $K=1$.
For intuitive explanation, consider the case in which $K=3$.
If we apply the same channel grouping rule used for even $K$ as shown in Fig. \ref{fig:odd_K}. (a), then messages are transmitted through a series of particular time indices $t_1$ to $t_3$ such that
\begin{equation}
\mathbf{H}_3[t_3]\mathbf{H}_2[t_2]\mathbf{H}_1[t_1]=\Big(\prod_{i=1}^3[\mathbf{\Gamma}]_{ii}\Big)\mathbf{U}\mathbf{V}^{\dagger},
\end{equation}
which is in general not a diagonal matrix.
Hence we apply the channel grouping rule as shown in Fig. \ref{fig:odd_K}. (b).
Then
\begin{equation}
\mathbf{H}_3[t_3]\mathbf{H}_2[t_2]\mathbf{H}_1[t_1]=\Big(\prod_{i=1}^3[\mathbf{\Gamma}]_{ii}\Big)\mathbf{I}_3
\end{equation}
and interference-free communication is possible.
However, the channel space partitioning of $\mathcal{U}_K$ used for even $K$ cannot guarantee that the probabilities of grouped channel subsets are the same.
To guarantee the same probabilities of grouped channel subsets, different partitioning method of $\mathcal{U}_K$ is proposed in this appendix.

We first introduce the quantization of a unitary matrix in the next two subsections and explain the channel space partitioning method and grouping rule.
Then we analyze its achievable DoF region.
We will use asymptotic relationships between two sequences $\{f(n)\}$ and $\{g(n)\}$.
We write $f(n){~}\dot{\leq}{~}g(n)$ if ${\lim\sup}_{n\to\infty}f(n)/g(n)\leq 1$ and $f(n)\doteq g(n)$ if ${\lim\inf}_{n\to\infty}f(n)/g(n)={\lim\sup}_{n\to\infty}f(n)/g(n)= 1$.

\subsection*{Quantization of Hypersphere}
Let $\mathcal{R}_m=\{\mathbf{a}\big|\|\mathbf{a}\|=1,\mathbf{a}\in\mathbb{R}^{m\times 1}\}$, where $m\geq 2$.
Then consider the quantization of  $\mathbf{r}\in\mathcal{R}_m$.
We first divide the set of angles defined in the hypersphere coordinates.
For $i\in\{2,\cdots,m-1\}$, define $\delta_i(k_1,\cdots,k_{i-1},\delta_1)=\frac{\delta_1}{\prod_{j=1}^{i-1}\cos(k_j\delta_j)}$, where $0<\delta_1<1$ and $k_j\in\mathbb{Z}$.
For $i\in\{1,\cdots,m-2\}$, define
\begin{eqnarray}
\!\!\!\!\!\!\!\!\!\!\!\!&&\mathcal{J}_i(k_1,\cdots,k_i)\nonumber\\
\!\!\!\!\!\!\!\!\!\!\!\!&&=[(k_i-1/2)\delta_i+\pi/2,(k_i+1/2)\delta_i+\pi/2),
\label{eq:quan_angle1}
\end{eqnarray}
where $|k_i|\leq \lfloor\frac{1}{\delta_i}\arccos(\delta_1^{1/(2(m-2))})\rfloor-1$.
Define
\begin{eqnarray}
\!\!\!\!\!\!\!\!\!\!\!\!\!\!\!\!\!\!&&\mathcal{J}_{m-1}(k_1,\cdots,k_{m-1})\nonumber\\
\!\!\!\!\!\!\!\!\!\!\!\!\!\!\!\!\!\!&&=[(k_{m-1}-1/2)\delta_{m-1}+\pi,(k_{m-1}+1/2)\delta_{m-1}+\pi),
\label{eq:quan_angle2}
\end{eqnarray}
where $k_{m-1}\in\mathbb{Z}$ and $|k_{m-1}|\leq \lfloor\frac{\pi}{\delta_{m-1}}\rfloor-1$.
From now on, $(k_1,\cdots,k_{m-1})$ will be assumed to be in the range specified above.

Let us define a quantized vector $\hat{\mathbf{r}}(k_1,\cdots,k_{m-1})=[\hat{r}_1,\cdots,\hat{r}_m]^T$, where
\begin{equation}
\hat{r}_i=\left(\prod_{j=1}^{i-1}\sin(k_j\delta_j+\pi/2)\right)\cos(k_i\delta_i+\pi/2)
\end{equation}
for $i=\{1,\cdots,m-2\}$,
\begin{equation}
\hat{r}_{m-1}=\left(\prod_{j=1}^{m-2}\sin(k_j\delta_j+\pi/2)\right)\cos(k_{m-1}\delta_{m-1}+\pi),
\end{equation}
and
\begin{equation}
\hat{r}_m=\left(\prod_{j=1}^{m-2}\sin(k_j\delta_j+\pi/2)\right)\sin(k_{m-1}\delta_{m-1}+\pi).
\end{equation}
Then denote $\mathcal{J}^{(m)}(\hat{\mathbf{r}}(k_1,\cdots,k_{m-1}))=\mathcal{J}_1(k_1)\times\mathcal{J}_2(k_1,k_2)\times\cdots\times \mathcal{J}_{m-1}(k_1,\cdots,k_{m-1})$.
For $\mathbf{r}\in\mathcal{R}_m$, the quantizer $\Delta_m$ is defined such that $\Delta_m(\mathbf{r})=\hat{\mathbf{r}}(k_1,\cdots,k_{m-1})$ if there exists $(k_1,\cdots,k_{m-1})$ satisfying $\mathbf{r}\in\mathcal{J}^{(m)}(\hat{\mathbf{r}}(k_1,\cdots,k_{m-1}))$, otherwise it declares an error.
We show that the following properties hold as the quantization interval $\delta_1$ converges to zero, which will be used to prove Lemma \ref{lemma:asymp_prop2}.

\begin{lemma} \label{lemma:asymp_prop1}
Suppose that $\delta_1$ is a function of $n$ such that $\delta_1(n)\to 0$ as $n\to \infty$.
Then the following properties hold:
\begin{enumerate}
\item If $\mathbf{r}$ is uniformly distributed over $\mathcal{R}_m$,
\begin{equation}
\Pr(\Delta_m(\mathbf{r})=\hat{\mathbf{r}}(k_1,\cdots,k_{m-1}))\doteq \frac{(\delta_1(n))^{m-1}\Gamma(m/2+1)}{m\pi^{m/2}}
\end{equation}
for any $(k_1,\cdots,k_{m-1})$, where $\Gamma(\cdot)$ denotes the Gamma function.
\item If $\mathbf{r}$ is uniformly distributed over $\mathcal{R}_m$, $\Pr(\lim_{n\to\infty}\|\mathbf{r}-\Delta_m(\mathbf{r})\|=0)=1$.
\end{enumerate}
\end{lemma}
\begin{proof}
Consider the first property.
Since $\mathbf{r}$ is uniformly distributed over $\mathcal{R}_m$, $\Pr(\mathbf{r}\in\mathcal{J}^{(m)}(\hat{\mathbf{r}}))$ is given as the volume of $\mathcal{J}^{(m)}(\hat{\mathbf{r}})$ divided by the volume of $\mathcal{R}_m$.
Then
\begin{eqnarray}
&&\operatorname{vol}(\mathcal{J}^{(m)}(\hat{\mathbf{r}}))\nonumber\\
&&\doteq\left(\prod_{i=1}^{m-1}\delta_i\right)\left(\prod_{i=1}^{m-2}(\sin(k_i\delta_i+\pi/2))^{m-1-i}\right)\nonumber\\
&&=\delta_1^{m-1},
\label{eq:prob_eq2}
\end{eqnarray}
where $\prod_{i=1}^{m-2}(\sin(k_i\delta_i+\pi/2))^{m-1-i}$ is the Jacobian of the volume of the hypersphere.
Therefore $\Pr(\Delta_m(\mathbf{r})=\hat{\mathbf{r}})\doteq \frac{\delta^{m-1}_{1}\Gamma(m/2+1)}{m\pi^{m/2}}$, where we use $\operatorname{vol}(\mathcal{R}_m)=\frac{m\pi^{m/2}}{\Gamma(m/2+1)}$ \cite{Lizhong:02}.

Consider the second property.
From the facts that
\begin{eqnarray}
&&\max_{k_i}\left\{k_i \delta_i\right\}\doteq \frac{\pi}{2}\mbox{ for }i\in\{1,\cdots,m-2\},\nonumber\\
&&\max_{k_{m-1}}\{k_{m-1}\delta_{m-1}\}\doteq\pi,
\end{eqnarray}
we have
\begin{equation}
\sum_{k_1,\cdots,k_{m-1}}\operatorname{vol}(\mathcal{J}^{(m)}(\hat{\mathbf{r}}(k_1,\cdots,k_{m-1})))\doteq\operatorname{vol}(\mathcal{R}_m)
\end{equation}
and, as a result, the outage probability tends to zero as $n$ increases.

Assume no outage from now on.
Let $\mathbf{r}=[r_1,\cdots,r_m]^T$ and $\Delta_m(\mathbf{r})=[\hat{r}_1,\cdots,\hat{r}_m]^T$.
First consider the case where $m\geq 3$.
From $|k_i|\leq \frac{1}{\delta_i}\arccos(\delta_1^{1/(2(m-2))})$ in (\ref{eq:quan_angle1}), $\cos(k_i\delta_i)\geq \delta_1^{1/(2(m-2))}$, where $i\in\{1,\cdots,m-2\}$.
By applying this inequality in the definition of $\delta_i$, we have $\delta_i\leq \delta_1^{1-(i-1)/(2(m-2))}$ for $i\in\{1,\cdots,m-1\}$.
From the hyperspherical coordinates, we also have
\begin{equation}
d r_1=\frac{\partial (\cos \phi_1)}{\partial\phi_1}d\phi_1,
\end{equation}
\begin{equation}
d r_i=\sum_{j=1}^i\frac{\partial (\sin \phi_1\cdots\sin \phi_{i-1}\cos \phi_i)}{\partial\phi_j}d\phi_j
\end{equation}
for $i\in\{2,\cdots,m-1\}$, and
\begin{equation}
 r_m=\sum_{j=1}^{m-1}\frac{\partial (\sin \phi_1\cdots\sin \phi_{m-2}\sin\phi_{n-1})}{\partial\phi_j}d\phi_j,
\end{equation}
which gives $|d r_i|\leq\sum_{j=1}^i|d\phi_j|$ for $i\in\{1,\cdots,m-1\}$ and $|d r_m|\leq\sum_{j=1}^{m-1}|d\phi_j|$.
Therefore $|r_i-\hat{r}_i|{~}\dot{\leq}{~}\sum_{j=1}^i\delta_j\leq (i-1)\delta_1^{1-(i-1)/(2(m-2))}\leq (m-1)\sqrt{\delta_1}$ for $i\in\{1,\cdots,m-1\}$.
Similarly, $|r_m-\hat{r}_m|{~}\dot{\leq}{~}\sum_{j=1}^{m-1}\delta_j\leq (m-1)\sqrt{\delta_1}$.
This means $\lim_{n\to\infty}\|\mathbf{r}-\Delta_m(\mathbf{r})\|=0$  $n\to\infty$.
The second property also holds for $m=2$ since $|r_1-\hat{r}_1|{~}\dot{\leq}{~}\delta_1$ and $|r_2-\hat{r}_2|{~}\dot{\leq}{~}\delta_1$ for this case.
In conclusion, Lemma \ref{lemma:asymp_prop1} holds.
\end{proof}

\subsection*{Quantization of Unitary Matrix}
From the hypersphere quantizer, we recursively quantize $\mathbf{U}\in\mathcal{U}_m$.
First consider $\mathbf{u}\in\mathbb{C}^{m\times 1}$ with $\|\mathbf{u}\|=1$.
Similar to $\mathcal{J}^{(m)}(\hat{\mathbf{r}})$, we can define $\mathcal{J}^{(2m)}(\hat{\mathbf{u}})$ in the $m$ dimensional complex space.
Then $\Delta_{2m}(\mathbf{u})=\hat{\mathbf{u}}$ quantizes $\mathbf{u}$ by treating it as a $2m$ dimensional real vector.
Let $\mathbf{u}=[u_1,\cdots,u_m]^T$.
For $i\in\{1,\cdots,m-1\}$, we define the $m\times m$ dimensional matrix $\mathbf{T}_i(\mathbf{u})$ such that $[\mathbf{T}_i(\mathbf{u})]_{11}=\frac{a^*_i}{\sqrt{|a_i|^2+|u_{i+1}|^2}}$, $[\mathbf{T}_i(\mathbf{u})]_{1(i+1)}=\frac{u^{*}_{i+1}}{\sqrt{|a_i|^2+|u_{i+1}|^2}}$, $[\mathbf{T}_i(\mathbf{u})]_{(i+1)1}=\frac{-u_{i+1}}{\sqrt{|a_i|^2+|u_{i+1}|^2}}$, $[\mathbf{T}_i(\mathbf{u})]_{(i+1)(i+1)}=\frac{a_{i}}{\sqrt{|a_i|^2+|u_{i+1}|^2}}$ and set the rest of diagonal elements as ones and the rest of off-diagonal elements as zeros.
Here, $a_1=u_1$ and $a_i\in\mathbb{R}_+$ is the first element of $(\prod_{j=1}^{i-1}\mathbf{T}_j(\mathbf{u}))\mathbf{u}$ for $i\in\{2,\cdots,m-1\}$.
Then define $\mathbf{T}(\mathbf{u})=\prod_{i=1}^{m-1}\mathbf{T}_i(\mathbf{u})$.
Note that $\mathbf{T}(\mathbf{u})\mathbf{u}=[1,\mathbf{0}_{1\times (m-1)}]^T$ because $\mathbf{T}(\mathbf{u})$ is a unitary matrix.
The quantizer $\Delta_{m\times m}:\mathbf{U}\to\hat{\mathbf{U}}$ is defined as follows:

\begin{itemize}
\item Set $\mathbf{U}'_1=\mathbf{U}$.

\item For $i\in\{1,\cdots,m\}$,\\
Let $\mathbf{u}'_i$ denote the first column vector of $\mathbf{U}'_i$.\\
Quantize $\mathbf{u}'_i$ such that $\Delta_{2(m-i+1)}(\mathbf{u}'_i)=\hat{\mathbf{u}}'_i$ if there exists $\hat{\mathbf{u}}'_i$ satisfying $\mathbf{u}'_i\in\mathcal{J}^{(2(m-i+1))}(\hat{\mathbf{u}}'_i)$, otherwise declare an error.\\
If $i\in\{1,\cdots,m-1\}$, set $\mathbf{U}'_{i+1}$ as $\mathbf{T}(\mathbf{u}'_i)\mathbf{U}'_i$ by removing the first column and the first row vectors of $\mathbf{T}(\mathbf{u}'_i)\mathbf{U}'_i$.\footnote{Here $\mathbf{T}(\mathbf{u}'_i)$ is the $(m-i+1)\times(m-i+1)$ dimensional matrix.}
That is,
\begin{equation}
\mathbf{T}(\mathbf{u}'_i)\mathbf{U}'_i=\left[
                             \begin{array}{cccc}
                               1 & \mathbf{0}_{1\times (m-i)} \\
                               \mathbf{0}_{(m-i)\times 1}&\mathbf{U}'_{i+1} \\
                             \end{array}
                           \right].
\label{eq:q_i}
\end{equation}

End.
\item Set $\Delta_{m\times m}(\mathbf{U})=\hat{\mathbf{U}}=[\hat{\mathbf{u}}_1,\cdots,\hat{\mathbf{u}}_m]$, where $\hat{\mathbf{u}}_1=\hat{\mathbf{u}}'_1$ and
\begin{equation}
\hat{\mathbf{u}}_i=\mathbf{T}^{\dagger}(\hat{\mathbf{u}}'_1)\cdots[0,[\mathbf{T}^{\dagger}(\hat{\mathbf{u}}'_{i-2})[0,[\mathbf{T}^{\dagger}(\hat{\mathbf{u}}'_{i-1})[0,{\hat{\mathbf{u}}_i}'^{T}]^T]^T]^T
\label{eq:u_delta}
\end{equation}
for $i\in\{2,\cdots,m\}$.
\end{itemize}

By using Lemma \ref{lemma:asymp_prop1}, we show that the following lemma holds as the quantization interval $\delta_1$ converges to zero.
These properties will be used to prove Theorem \ref{th:dof_K_K} for odd $K$.

\begin{lemma} \label{lemma:asymp_prop2}
Suppose that $\delta_1$ is a function of $n$ such that $\delta_1(n)\to 0$ as $n\to \infty$.
Then the following properties hold:
\begin{enumerate}
\item If $\mathbf{U}$ is uniformly distributed over $\mathcal{U}_m$,
\begin{eqnarray}
&&\!\!\!\!\!\!\!\!\!\!\!\!\!\!\!\!\!\Pr(\Delta_{m\times m}(\mathbf{U})=\Delta_{m\times m}(\mathbf{V}))\nonumber\\
&&\!\!\!\!\!\!\!\!\!\!\!\!\!\!\!\!\!\doteq\left(\prod_{i=1}^m(\delta_1(n))^{2(m-i)+1}\right)\left(\prod_{i=1}^m\frac{(i-1)!}{2\pi^i}\right)
\end{eqnarray}
for any $\mathbf{V}\in\mathcal{U}_m$ such that $\Delta_{m\times m}(\mathbf{V})$ exists.
\item If $\mathbf{U}$ is uniformly distributed over $\mathcal{U}_m$,
\begin{equation}
\Pr(\lim_{n\to\infty}\|\mathbf{U}-\Delta_{m\times m}(\mathbf{U})\|_F=0)=1.
\end{equation}
\end{enumerate}
\end{lemma}
\begin{proof}
Consider the first property.
From the first property of Lemma \ref{lemma:asymp_prop1},
\begin{eqnarray}
\!\!\!\Pr(\Delta_{m\times m}(\mathbf{U})\!\!\!\!\!\!\!\!\!\!&&=\Delta_{m\times m}(\mathbf{V}))\nonumber\\
\!\!\!&&\doteq\prod_{i=1}^m \frac{\delta_1^{2(m-i)+1}\Gamma(m-i+2)}{2(m-i+1)\pi^{m-i+1}}\nonumber\\
\!\!\!&&=\left(\prod_{i=1}^m\delta_1^{2(m-i)+1}\right)\left(\prod_{i=1}^m\frac{(i-1)!}{2\pi^i}\right).
\end{eqnarray}
Let $\mathbf{U}=[\mathbf{u}_1,\cdots,\mathbf{u}_m]$ and $\Delta_{m\times m}(\mathbf{U})=[\hat{\mathbf{u}}_1,\cdots,\hat{\mathbf{u}}_m]$.
Since $\Delta_{2m}(\mathbf{u}_1)=\hat{\mathbf{u}}_1$, from the second property of Lemma \ref{lemma:asymp_prop1}, $\hat{\mathbf{u}}_1\to\mathbf{u}_1$ with probability one as $n\to\infty$.
Then consider $\mathbf{u}_2$ and $\hat{\mathbf{u}}_2$ satisfying $\Delta_{2(m-1)}(\mathbf{u}'_2)=\hat{\mathbf{u}}'_2$, where $\mathbf{T}(\mathbf{u}_1)\mathbf{u}_2=[0,\mathbf{u}_2'^T]^T$ and $\mathbf{T}(\hat{\mathbf{u}}_1)\hat{\mathbf{u}}_2=[0,\hat{\mathbf{u}}_2'^T]^T$.
Here, $\mathbf{T}(\mathbf{u}_1)\mathbf{u}_2=[0,\mathbf{u}_2'^T]^T$ is obtained from (\ref{eq:q_i}) and $\mathbf{u}'_1=\mathbf{u}_1$ and $\mathbf{T}(\hat{\mathbf{u}}_1)\hat{\mathbf{u}}_2=[0,\hat{\mathbf{u}}_2'^T]^T$ is obtained from (\ref{eq:u_delta}) and $\hat{\mathbf{u}}'_1=\hat{\mathbf{u}}_1$.
Since $\mathbf{T}(\cdot)$ is a continuous function and $\hat{\mathbf{u}}_1\to\mathbf{u}_1$ with probability one, $\mathbf{T}(\hat{\mathbf{u}}_1)\to\mathbf{T}(\mathbf{u}_1)$ with probability one as $n\to\infty$.
From the second property of Lemma \ref{lemma:asymp_prop1}, $\hat{\mathbf{u}}'_2\to\mathbf{u}'_2$ with probability one.
By using these two facts, we have
\begin{equation}
\hat{\mathbf{u}}_2=\mathbf{T}(\hat{\mathbf{u}}_1)^{\dagger}[0,\hat{\mathbf{u}}_2'^T]^T\to\mathbf{T}(\mathbf{u}_1)^{\dagger}[0,\mathbf{u}_2'^T]^T=\mathbf{u}_2
\end{equation}
with probability one as $n\to\infty$.
By applying the same analysis recursively, we can show that $\Pr(\lim_{n\to\infty}\|\mathbf{U}-\Delta_{m\times m}(\mathbf{U})\|_F=0)=1$.
In conclusion, Lemma \ref{lemma:asymp_prop2} holds.
\end{proof}

\subsection*{DoF Region for Odd $K$}
In this subsection, we prove that Theorem \ref{th:dof_K_K} holds for odd $K$.
We will briefly describe the differences from the proposed scheme for even $K$ in Section \ref{sec:dof_K_K}.
First, we quantize unitary matrices by using the quantizer described in the previous subsection.
Based on the new quantizer, we can define $\mathcal{Q}_{\delta}$ and $\mathcal{Q}(\mathbf{U}_{\delta})$ as in Section \ref{sec:dof_K_K}, where, we again use $\mathcal{Q}_{\delta}$ and $\mathcal{Q}(\mathbf{U}_{\delta})$ notations for notational convenience.
Specifically, $\mathcal{Q}_{\delta}$ is the set of all quantization points of $\Delta_{K\times K}(\mathbf{A})$ for $\mathbf{A}\in\mathcal{U}_K$ and $\mathcal{Q}(\mathbf{U}_{\delta})$ is the set of all $\mathbf{A}\in\mathcal{U}_K$ satisfying $\Delta_{K\times K}(\mathbf{A})=\mathbf{U}_{\delta}$, where $\mathbf{U}_{\delta}\in\mathcal{Q}_{\delta}$.
Then we can modify Lemma \ref{lemma:eq_prob} as follows:
Suppose that $\delta_1$ is a function of $n$ such that $\delta_1(n)\to 0$ as $n\to \infty$.
For $\mathbf{\Sigma}_{\delta}^{(1)}\in\mathcal{I}_{\delta}$ and $\mathbf{\Sigma}_{\delta}^{(2)}\in\mathcal{I}_{\delta}$, if there exists a permutation matrix $\mathbf{\Gamma}$ such that $\mathbf{\Sigma}_{\delta}^{(2)}=\mathbf{\Gamma}^T\mathbf{\Sigma}_{\delta}^{(1)}\mathbf{\Gamma}$, then
\begin{eqnarray}
&&\Pr(S(\mathbf{H}_m[t])\in\mathcal{S}(\mathbf{U}^{(1)}_{\delta},\mathbf{\Sigma}^{(1)}_{\delta},\mathbf{V}^{(1)}_{\delta}))\nonumber\\
&&\overset{(a)}{=}\Pr(\mathbf{U}_m[t]\in\mathcal{Q}(\mathbf{U}^{(1)}_{\delta}))\Pr(\mathbf{\Sigma}_m[t]\in\mathcal{I}(\mathbf{\Sigma}^{(1)}_{\delta}))\nonumber\\
&&{~~}\cdot\Pr(\mathbf{V}_m[t]\in\mathcal{Q}(\mathbf{V}^{(1)}_{\delta}))\nonumber\\
&&\overset{(b)}{\doteq}\Pr(\mathbf{U}_m[t]\in\mathcal{Q}(\mathbf{U}^{(2)}_{\delta}))\Pr(\mathbf{\Sigma}_m[t]\in\mathcal{I}(\mathbf{\Sigma}^{(2)}_{\delta}))\nonumber\\
&&{~~}\cdot\Pr(\mathbf{V}_m[t]\in\mathcal{Q}(\mathbf{V}^{(2)}_{\delta}))\nonumber\\
&&{=}\Pr(S(\mathbf{H}_m[t])\in\mathcal{S}(\mathbf{U}^{(2)}_{\delta},\mathbf{\Sigma}^{(2)}_{\delta},\mathbf{V}^{(2)}_{\delta}))
\label{eq:eq_prob_asym}
\end{eqnarray}
for all $\mathbf{U}^{(1)}_{\delta},\mathbf{U}^{(2)}_{\delta},\mathbf{V}^{(1)}_{\delta},\mathbf{V}^{(2)}_{\delta}\in\mathcal{Q}_{\delta}$, where  $S(\mathbf{H}_m[t])$ denotes $(\mathbf{U}_m[t],\mathbf{\Sigma}_m[t],\mathbf{V}_m[t])$.
Here, $(a)$ holds from Lemma \ref{lemma:product_pdf},
$(b)$ holds since $\mathbf{U}_m[t]$ and $\mathbf{V}_m[t]$ are uniformly distributed over $\mathcal{U}_K$ (Lemma \ref{lemma:product_pdf}) and from the first property of Lemma \ref{lemma:asymp_prop2}.

Second, we apply the different channel grouping by modifying the relaying of the proposed scheme in Section \ref{sec:dof_K_K} as follows:
\begin{itemize}
\item (Relaying for $m=\{2,\cdots,K\}$)
\\
For all $(\mathbf{U}_{\delta},\mathbf{\Sigma}_{\delta},\mathbf{V}_{\delta})\in\mathcal{Q}_{\delta}\times\mathcal{I}_{\delta}\times\mathcal{Q}_{\delta}$, the nodes in the $m^{\operatorname{th}}$ layer amplify and forward their received signals that are received during $\cup_{\mathbf{U}'_{\delta}\in\mathcal{Q}_{\delta}}\mathcal{T}_{m-1}(\mathbf{V}_{\delta},\mathbf{P}^T\mathbf{\Sigma}_{\delta}\mathbf{P},\mathbf{U}'_{\delta})$ using $N(\mathbf{U}_{\delta},\mathbf{\Sigma}_{\delta},\mathbf{V}_{\delta})$ time indices in $\mathcal{T}_m(\mathbf{U}_{\delta},\mathbf{\Sigma}_{\delta},\mathbf{V}_{\delta})$.
If $m=K$, it is also satisfied that these signals are received during $\cup_{\mathbf{V}'_{\delta}\in\mathcal{Q}_{\delta}}\mathcal{T}_1(\mathbf{V}'_{\delta},(\mathbf{P}^T)^{K-1}\mathbf{\Sigma}_{\delta}\mathbf{P}^{K-1},\mathbf{U}_{\delta})$ at the first hop.
\end{itemize}

Similar to (\ref{eq:paired_channels}), messages are transmitted through $t_1$ to $t_K$ such that
\begin{equation}
S(\mathbf{H}_m[t_m])\in\mathcal{S}(\mathbf{U}_{\delta,m},\mathbf{P}^{m-1}\mathbf{\Sigma}_{\delta}(\mathbf{P}^T)^{m-1},\mathbf{V}_{\delta,m}),
\end{equation}
where $\mathbf{U}_{\delta,1}=\mathbf{V}_{\delta,2},\mathbf{U}_{\delta,2}=\mathbf{V}_{\delta,3},\cdots,\mathbf{U}_{\delta,K-1}=\mathbf{V}_{\delta,K}$, and $\mathbf{V}_{\delta,1}=\mathbf{U}_{\delta,K}$.
Hence interference-free communication is again possible if the quantization intervals $\delta$ and $\delta_1$ converge to zero.

Then almost the same proof for even $K$ in Theorem \ref{th:dof_K_K} can be applied for odd $K$.
We briefly explain the differences.
Let $P_{\min}(\mathbf{\Sigma}_{\delta})=\min_{\mathbf{U}_{\delta}\in\mathcal{Q}_{\delta},\mathbf{V}_{\delta}\in\mathcal{Q}_{\delta},\mathbf{\Gamma}\in \mathcal{Q}_{\Gamma}} P(\mathbf{U}_{\delta},\mathbf{\Gamma}^T\mathbf{\Sigma}_{\delta}\mathbf{\Gamma},\mathbf{V}_{\delta})$, where $\mathcal{Q}_{\Gamma}$ denotes the set of all permutation matrices.
Set $N(\mathbf{U}_{\delta},\mathbf{\Sigma}_{\delta},\mathbf{V}_{\delta})=\max\{\lfloor n_BP_{\min}(\mathbf{\Sigma}_{\delta})-\epsilon,0 \rfloor\}$.
Since $N(\mathbf{U}_{\delta},\mathbf{\Sigma}_{\delta},\mathbf{V}_{\delta})$ is the same for all $\mathbf{U}_{\delta}\in\mathcal{Q}_{\delta},\mathbf{V}_{\delta}\in\mathcal{Q}_{\delta}$, and $\mathbf{\Gamma}\in \mathcal{Q}_{\Gamma}$, every transmit signal can be delivered to the final destinations if $E_{1,i}$ does not occur.
Because of the different quantization of unitary matrices, $\operatorname{card}(\mathcal{Q}_{\delta})\leq (\frac{2\pi}{\delta_1})^{2K^2}$ since the number of points of $k_i$ used in (\ref{eq:quan_angle1}) and (\ref{eq:quan_angle2}) is less than or equal to $\frac{2\pi}{\delta_1}$.
Then by setting $\delta_1=\alpha^{-1}$, we have $\operatorname{card}(\mathcal{Q}_{\delta})\leq (2\pi)^{2K^2}\alpha^{2K^2}$.
For even $K$, $\operatorname{card}(\mathcal{Q}_{\delta})\leq (2\alpha+1)^{2K^2}\leq3^{2K^2}\alpha^{2K^2}$ was used.
Hence $\delta=n_B^{-1/(32K^2)}$, $\alpha=n_B^{1/(16K^2)}$, and $\epsilon=n_B^{-1/3}$ again satisfy the conditions (\ref{eq:condition1}) to (\ref{eq:condition4}).
Lastly, as $n_B\to \infty$ (equivalently, $\delta_1\to 0$ and $\delta\to 0$), $P(\mathbf{U}_{\delta},\mathbf{\Sigma}_{\delta},\mathbf{V}_{\delta})$ is asymptotically the same for all $\mathbf{U}_{\delta}\in\mathcal{Q}_{\delta},\mathbf{V}_{\delta}\in\mathcal{Q}_{\delta}$, and $\mathbf{\Gamma}\in \mathcal{Q}_{\Gamma}$ and the quantization errors converge to zero with probability one, where we use the first and second properties of Lemma \ref{lemma:asymp_prop2}.
Therefore, we can derive the same equation as in (\ref{eq:final_rate}).
In conclusion, Theorem \ref{th:dof_K_K} holds for odd $K$.


\begin{biographynophoto}
{Sang-Woon Jeon}(S'07) received the B.S. and the M.S. degrees in Electrical Engineering from Yonsei University, Seoul, Korea in 2003 and 2006, respectively.
He is currently working toward the Ph.D. degree at KAIST, Daejeon, Korea.
His research interests include network information theory and its application to communication systems.
\end{biographynophoto}

\begin{biographynophoto}
{Sae-Young Chung}(S'89--M'00--SM'07) received the B.S. and the M.S. degrees in Electrical Engineering from Seoul National University in 1990 and 1992, respectively. He received the Ph.D. degree in Electrical Engineering and Computer Science at MIT in 2000. From June to August 1998 and from June to August 1999, he was with Lucent technologies. From September 2000 to December 2004, he was with Airvana, Inc., where he conducted research on the third generation wireless communications. Since January 2005, he has been with KAIST, where he is now an Associate Professor in the Department of Electrical Engineering. He is currently serving as an Editor of the IEEE transactions on Communications. He served as a guest editor of JCN, EURASIP journal, and Telecommunications Review. He is serving as a TPC co-chair of the 2014 IEEE International Symposium on Information Theory (ISIT) to be held in Hawaii. He also served as a TPC co-chair of WiOpt 2009. His research interests include network information theory, coding theory, and their applications to wireless communications. He is a senior member of the IEEE.
\end{biographynophoto}

\begin{biographynophoto}
{Syed Ali Jafar} (S'99--M'04--SM'09) received the B. Tech. degree in Electrical Engineering from
the Indian Institute of Technology (IIT), Delhi, India in 1997, the M.S.
degree in Electrical Engineering from California Institute of Technology
(Caltech) , Pasadena USA in 1999, and the Ph.D. degree in Electrical
Engineering from Stanford University, Stanford, CA USA in 2003. His
industry experience includes positions at Lucent Bell Labs , Qualcomm Inc.
and Hughes Software Systems. He is currently an Associate Professor in the
Department of Electrical Engineering and Computer Science at the
University of California Irvine, Irvine, CA USA. His research interests
include multiuser information theory and wireless communications.

Dr. Jafar received the NSF CAREER award in 2006, the ONR Young
Investigator Award in 2008, the IEEE Information Theory Society paper
award in 2009 and the UCI Engineering School Fariborz Maseeh Outstanding
Research Award in 2010.  He received the UC Irvine Engineering Faculty of
the Year award in 2006 and the UC Irvine EECS Professor of the Year Award
in 2009, for excellence in teaching. He served as Associate Editor for
IEEE Transactions on Communications 2004-2009, for IEEE Communications
Letters 2008-2009 and is currently serving as Associate Editor for IEEE
Transactions on Information Theory.
\end{biographynophoto}

\end{document}